\documentclass[useAMS]{mn2e}

\usepackage{epsfig}

\newcommand{\be}{\begin{eqnarray}}
\newcommand{\ee}{\end{eqnarray}}
\def\lsim{\,\lower2truept\hbox{${< \atop\hbox{\raise4truept\hbox{$\sim$}}}$}\,}
\def\gsim{\,\lower2truept\hbox{${> \atop\hbox{\raise4truept\hbox{$\sim$}}}$}\,}
\title[The first low-mass stars]{The first low-mass stars: critical metallicity or dust-to-gas ratio?}
\author[Raffaella Schneider, Kazuyuki Omukai, Simone Bianchi, Rosa Valiante]{Raffaella Schneider$^{1}$\thanks{E-mail:
raffaella.schneider@oa-roma.inaf.it}, Kazuyuki Omukai$^{2}$, Simone Bianchi$^{3}$, Rosa Valiante$^{3}$\\
$^{1}$ INAF/Osservatorio Astronomico di Roma, Via di Frascati 33, 00040 Monteporzio, Roma, Italy\\
$^{2}$ Department of Physics, Kyoto University, Kyoto 606-8502, Japan\\
$^{3}$ INAF/Osservatorio Astrofisico di Arcetri, Largo Enrico Fermi 5, 50125 Firenze, Italy} 
\begin{document}

\date{September 2011}

\pagerange{\pageref{firstpage}--\pageref{lastpage}} \pubyear{2011}

\maketitle

\label{firstpage}

\begin{abstract}
We explore the minimal conditions which enable the formation of metal--enriched solar and sub--solar
mass stars. Using a one--zone semi-analytical model, we accurately follow the chemical and
thermal evolution of the gas with the aim of understanding how the initial metal and dust content 
alters the cooling and fragmentation properties, hence the characteristic stellar mass.
We find that in the absence of dust grains, gas fragmentation occurs at densities 
$n_{\rm H} \sim [10^4-10^5]$~cm$^{-3}$ when the metallicity exceeds $Z \sim 10^{-4} Z_{\odot}$. 
The resulting fragmentation masses are $\ge 10 M_{\odot}$.
The inclusion of Fe and Si cooling does not affect the thermal evolution as this is dominated by
molecular (mostly OH, H$_2$O and CO) cooling even for metallicities as large as
$Z = 10^{-2} Z_{\odot}$.
The presence of dust is the key driver for the formation of low--mass stars.  
We focus on three representative core--collapse supernova (SN) progenitors
(a $Z=0$ star with $20 M_{\odot}$ and two $Z=10^{-4} Z_{\odot}$ stars
with $20 M_{\odot}$ and $35 M_{\odot}$), 
and consider the effects of reverse shocks of increasing strength: these
reduce the depletion factors, $f_{\rm dep}=M_{\rm dust}/(M_{\rm dust}+M_{\rm met})$, 
alter the shape of the grain size distribution function and modify the relative abundances of grain species and
of metal species in the gas phase. 
We find that the lowest metallicity at which fragmentation occurs is $Z=10^{-6} Z_{\odot}$ 
for gas pre--enriched by the explosion of a $20 M_{\odot}$ primordial SN ($f_{\rm dep} \ge 0.22$) and/or by a $35 M_{\odot}$,
$Z=10^{-4} Z_{\odot}$ SN ($f_{\rm dep} \ge 0.26$);
it is $\sim$ 1 dex larger, when the gas is pre--enriched by a $Z = 10^{-4} Z_{\odot}$, 20~$M_{\odot}$ SN ($f_{\rm dep} \ge 0.04$).
Cloud fragmentation depends on the depletion factor and it is suppressed when the
reverse shock leads to a too large destruction of dust grains.
These features are all consistent with the existence of a minimum dust--to--gas ratio, 
${\cal D}_{\rm cr}$, above which fragmentation is activated. We derive a simple analytic
expression for ${\cal D}_{\rm cr}$ which depends on the
total grain cross--section per unit mass of dust; for grain
composition and properties explored in the present study, 
${\cal D}_{\rm cr} = [2.6 - 6.3] \times 10^{-9}$.
When the dust--to--gas ratio of star forming clouds exceeds this value,
the fragmentation masses range between $0.01 M_{\odot}$ and $1 M_{\odot}$,
thus enabling the formation of the first low--mass stars.
\end{abstract}


\section{Introduction}

The formation of the first low--mass and long--lived stars marks an important
step in cosmic evolution. Despite that a large fraction of the Galactic Halo has now
been surveyed (Beers \& Christlieb 2005; Helmi 2008), no evidence for the existence of primordial low--mass stars
has been found to date. The most primitive star of the Galactic Halo 
has a [Fe/H] = -4.99 and, contrary to other extremely iron--poor stars, shows no carbon, 
nitrogen or oxygen enhancement, resulting in a total metallicity of $Z = 4.5 \times 10^{-5} Z_{\odot}$ (Caffau et al. 2011).

This lends support to the idea that the formation of
solar or sub--solar mass stars\footnote{For WMAP7 cosmological parameters, the
current Hubble time is $t_{\rm H} = 13.75$~Gyr (Jarosik et al. 2011). Depending on the stellar initial metallicity,
this corresponds to the lifetime of stars with masses in the range $[0.7 - 0.9]M_{\odot}$
(Raiteri et al. 1996)} requires some minimal conditions which 
do not rely on carbon and oxygen line--cooling (Frebel, Johnson, and Bromm 2007) 
and that can operate at very low but non-zero metallicity (Schneider et al. 2003, 2006).

This empirical evidence is supported by a solid theoretical argument: the micro--physics
of the H$_2$ molecule, the only available coolant in $\sim 10^6 M_{\odot}$ mini--halos at
$z \sim 20-30$, sets a characteristic thermodynamic state where gravitational fragmentation
of star--forming clouds is halted, which corresponds to a minimum Jeans mass of $\sim 500 M_{\odot}$ (see Bromm et al. 2009 and references therein).
Simulations starting from cosmological initial conditions have confirmed that vigorous fragmentation
of collapsing clouds does not occur (Abel, Bryan \& Norman 2002; Bromm, Coppi \& Larson 2002;
Yoshida et al. 2006; Yoshida, Omukai \& Hernquist 2008), although there have been evidences
for the formation of massive binary systems (Turk, Abel, O'Shea 2009).  

The subsequent phase of gas accretion onto the central proto--stellar core is subject to 
radiative feedback effects from the growing protostar which leads to a plausible but 
uncertain range of final stellar masses, $[30 - 100]~M_{\odot}$ (Omukai \& Palla 2003; Tan \& McKee 2004; 
Bromm \& Loeb 2004; Yoshida et al. 2008; McKee \& Tan 2008; 
Hosokawa \& Omukai 2009; Ohkubo et al. 2009). The accretion disk may become susceptible to fragmentation,
forming groups of protostars (Machida et al. 2008; Stacy, Greif \& Bromm 2010; Clark et al. 2011; 
Greif et al. 2011), but current simulations are not able to follow the subsequent interactions
of group members with the surrounding gas cloud.

Rotation, turbulence and magnetic fields are known to play an important role in present--day 
star formation and their importance in the first star forming regions at high redshift still 
needs to be fully assessed (see e. g. Schleicher et al. 2009; Clark et al. 2011b and references therein). 
Whether or not these additional physical processes might enable the formation of low--mass 
Pop III stars remains an open question. 

Here we follow a different approach: guided by the observed constancy of the stellar characteristic mass
in the local Universe, $M_{\rm ch} \sim 0.3 M_{\odot}$ 
(for recent reviews see Elmegreen 2009; Bastian, Covey \& Meyer 2010), which has been interpreted 
to originate from the minimum Jeans mass imprinted during gravitational fragmentation 
(Li, Klessen \& McLow 2003; Larson 2005; Jappsen et al. 2005; Elmegreen, Klessen \& Wilson 2008),
we explore the most extreme environmental conditions where low--mass star formation can occur
in the first mini--halos at high redshifts. 

The transition in the characteristic stellar mass, often referred to as Pop III/II transition,
has been explored by many authors. On the relatively small scale of a single star forming cloud,
semi-analytic and numerical studies have emphasized the role of metals and dust. 

The effects of metal--line cooling, predominantly by O, C, Si and Fe, are relevant at relatively
low densities, $\le [10^4 - 10^5]$~cm$^{-3}$, 
before and around the onset of NLTE-LTE transition of the corresponding level populations.
When the initial metallicity of the star forming cloud exceeds a critical value of $Z_{\rm cr} \sim 
[10^{-4} - 10^{-3}] Z_{\odot}$,
the minimum Jeans mass achieved by gravitational fragmentation is found to vary in the range $[10 - 100] M_{\odot}$,
depending on the presence/absence of molecular coolants in the chemical network and on the number of
heating/cooling processes included in the modelling (Omukai 2000; Bromm et al. 2001; Schneider et al. 2002; 
Bromm \& Loeb 2003; Santoro \& Shull 2006; Jappsen et al. 2007, 2009; 
Smith, Sigurdsson \& Abel 2008; Hocuk \& Spaans 2010). 
In this case, the formation of sub--solar or solar
mass stars, which may survive to the present time, is determined during the subsequent evolutionary phases, by gas accretion
within each clump and dynamical interactions within the cluster (Machida et al. 2009; Omukai, Hosokawa \& Yoshida 2010). 

As in present--day conditions, thermal emission by collisionally excited dust grains is an efficient 
cooling channel at high
densities, $\ge [10^{12} - 10^{14}]$~cm$^{-3}$.  As a result, the typical Jeans masses which can be
achieved by gravitational fragmentation are much smaller, $[0.01-1]~M_{\odot}$ 
(Schneider et al. 2003, 2006; Omukai et al. 2005; Tsuribe \& Omukai 2006; 
Clark, Glover \& Klessen 2008; Omukai et al. 2010; Dopcke et al. 2011). 
Thus, efficient dust--cooling leads to a transition in the characteristic mass by several orders of magnitude
with respect to the primordial case. The metallicity at which this transition takes place depends on the
dust depletion factor, defined as,
\begin{equation}
f_{\rm dep} = \frac{M_{\rm dust}}{M_{\rm dust} + M_{\rm met}},
\end{equation}
\noindent
where $M_{\rm dust}$ and $M_{\rm met}$ are the dust and gas-phase
metals masses, respectively. 

In the model presented in Omukai et al. (2005), where we assumed present-day ISM composition and grain properties,
$f_{\rm dep} = 0.49$ (Pollack et al. 1994) and the onset of dust-induced cooling and fragmentation
leading to the formation of $\sim 1 M_\odot$ fragments starts to be effective
when $10^{-6} Z_{\odot} < Z_{\rm cr} \le 10^{-5} Z_{\odot}$.
This result has been confirmed by numerical simulations 
(Tsuribe \& Omukai 2006, 2008; Clark et al. 2008; Omukai et al. 2010; Dopcke et al. 2011).

Since the properties of dust grains at high redshifts might be different with respect to
those observed locally, in Schneider et al. (2006) we have investigated the 
dust--driven transition under the hypothesis that the collapsing star-forming cloud
had been pre-enriched by metal-free core-collapse supernova (with a progenitor of
$22 M_{\odot}$) and by a pair-instability supernova (with a
progenitor of $195 M_{\odot}$). The adopted metal composition and dust-grain
properties were computed according to the model of dust nucleation in supernova
ejecta developed by Todini \& Ferrara (2001) and by Schneider, Salvaterra \& Ferrara (2003).
The depletion factors were $f_{\rm dep} = 0.23 $ (for the $22 M_{\odot}$ progenitor) and
$f_{\rm dep} = 0.65$ (for the $195 M_{\odot}$) and low-mass fragmentation was found to
occur already at $Z_{\rm cr} = 10^{-6} Z_{\odot}$.

An important confirmation of a Pop III/II transition driven by dust cooling has 
been recently provided by Caffau et al. (2011), who have reported the observation
of a Galactic Halo star with $Z = 4.5 \times 10^{-5} Z_{\odot}$. 
Since this paper
has appeared at the time of submission of the present work, we defer a specific analysis
of this particular star to a future study.

Here we intend to investigate the effects of varying the initial $f_{\rm dep}$ 
on the gravitational collapse and fragmentation properties of pre--enriched star forming
clouds. The questions we intend to address are: {\it (i)} what is the minimum fragment mass
that can be achieved by metal--line cooling? and {\it (ii)} What is the minimum dust depletion 
factor to activate dust-induced cooling?
To answer these questions, we first calculate the thermal evolution of collapsing
gas clouds that cool only through molecular (H$_2$, HD, H$_2$O, CO, OH) 
and metal (CI, CII, OI, SiII, FeII) line cooling. Then, 
we use the model developed by Bianchi \& Schneider (2007) to compute the mass, 
composition and size distribution of dust formed in the ejecta of a few representative 
SN progenitors. Starting from these initial conditions, we take into account the 
partial destruction of the newly formed dust by the SN reverse shock. In particular,
we consider reverse shocks of increasing strength, assuming increasingly large circumstellar
densities. 
Following the models by Bianchi \& Schneider (2007) 
we can control the reduction of the depletion factor (more than 80\% of the initial dust mass is destroyed)
in a self--consistent way and explore the implications
for the fragmentation properties of pre--enriched star forming clouds.

The plan of the paper is as follows: 
in Section 2 we present the model; in Section 3 we 
discuss the thermal evolution driven by line--cooling
and the resulting fragmentation properties.
In Section 4 we describe the adopted model grid of core--collapse SN
progenitors and the associated dust properties. 
In Section 5 we present the thermal evolution and fragmentation of
clouds pre--enriched by these explosions. 
In Section 6 we discuss the results and, finally, in section 7 we 
summarize our main conclusions.

\section{The model}
\label{sec:model}
The thermal and chemical evolution of collapsing gas clouds is 
investigated using the model presented in Omukai (2000) and
in Omukai et al. (2005) with CO, H$_2$O and
OH cooling rates recently modified as in Omukai et al. (2010).
As a new feature of the code, we have included the contribution
of Fe and Si to metal--line cooling.
Here we briefly summarize the features of the model which are
relevant for the purposes of the present study and we refer
the interested reader to the original papers for a thorough
description of the code. 

The temperature evolution is computed by solving the energy equation,

\begin{equation}
\frac{de}{dt}=-p  \,\frac{d}{dt} \,\frac{1}{\rho} - \Lambda_{\rm net}
\label{eq:energy}
\end{equation}
\noindent
where the pressure, $p$,  and the specific thermal energy, $e$, are
given by,

\be
p = \frac{\rho  \,k  \,T}{\mu  \,m_{\rm H}}
\label{eq:gas}
\ee

\be
e=\frac{1}{\gamma-1} \frac{k \,T}{\mu  \,m_{\rm H}}
\ee
\noindent
and $\rho$ is the central density, $T$ is the temperature, $\gamma$ is the adiabatic exponent,
$\mu$ is the mean molecular weight\footnote{The mean molecular weight is 1.23 for a fully atomic gas
and 2.29 for a fully molecular gas.} and $m_{\rm H}$ is the mass of hydrogen nuclei.
 
The terms on the right-hand side of the energy
equation are the compressional heating rate,

\be
\frac{d\rho}{dt} = \frac{\rho}{t_{\rm col}}  
\qquad \mbox{with} \qquad t_{\rm col} = \frac{1}{\sqrt{1-f_{\rm p}}} \sqrt{\frac{3 \pi}{32 G \rho}},
\label{eq:compression}
\ee

\noindent
where $f_{\rm p}$ is the ratio between pressure gradient to gravity (see Omukai et al. 2005 
for the expression) 
and the net cooling rate, $\Lambda_{\rm net}$, which consists
of three components,

\begin{equation}
\Lambda_{\rm net} = \Lambda_{\rm line} + \Lambda_{\rm cont} + \Lambda_{\rm chem}.
\label{eq:cool}
\end{equation}
\noindent
The first component, $\Lambda_{\rm line}$, represents the cooling rate due to the emission of line radiation,
which includes molecular line emission of H$_2$, HD, OH, H$_2$O and CO (Omukai et al. 2010) 
and atomic fine-structure line emission. In addition to CI, CII and OI (Omukai et al. 2005), here we have
considered the contribution of Fe and Si lines (see section \ref{sec:fesicool} below). 
The second component, $\Lambda_{\rm cont}$, represents the cooling rate due to
the emission of continuum radiation, which includes thermal emission by dust grains and H$_2$ collision-induced
emission. The last term, $\Lambda_{\rm chem}$, indicates the cooling/heating rate due to chemical reactions.
When the gas cloud is optically thick to a specific cooling radiation, the cooling rate is correspondingly
reduced by multiplying by the photon escape probability (see Omukai 2000). 

We do not consider the effects of the Cosmic Microwave Background (CMB)
radiation field, which provides a redshift dependent effective temperature floor 
(Clarke \& Bromm 2003; Omukai et al. 2005). At $z \ge 10-15$, the evolution
of star--forming clouds can be significantly affected but only at relatively 
large metallicities, $Z \ge 10^{-2} Z_{\odot}$ (Smith et al. 2009; 
Schneider \& Omukai 2010). We will return to this point in section 6. 

The effects of dust grains on the 
thermal evolution are due to {\it (i)} dust thermal emission and {\it (ii)} the increased
formation rate of H$_2$ on its surface and {\it (iii)} the heating associated to the latter process. 
The treatment of these physical processes is the same as in Schneider et al. (2006).
Here we report only the fundamental equations which regulate dust thermal emission and refer the interested reader to 
Schneider et al. (2006) for further details.

\begin{figure}
\center{\epsfig{figure=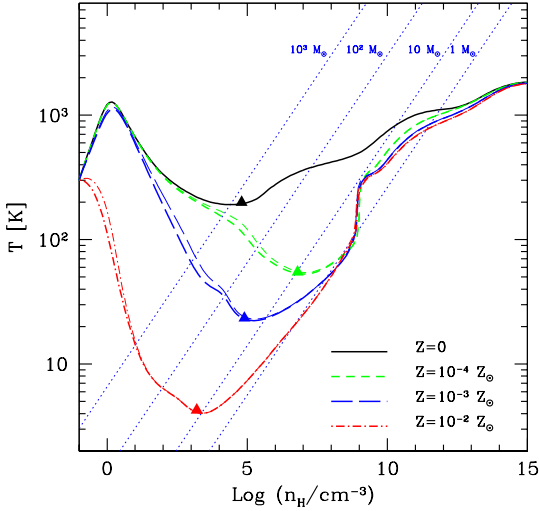, height=8.5cm}}
\caption{Thermal evolution of collapsing clouds with initial
metallicities $Z=0, 10^{-4} Z_{\odot}, 10^{-3} Z_{\odot}, 10^{-2} Z_{\odot}$,
from top to bottom. For each of the $Z>0$ models, two lines are shown: the thin one
represents the prediction of the Omukai et al. (2005) model; the thick
one shows the effects of including [FeII] and [SiII] cooling. 
The diagonal dotted lines indicate locii of constant Jeans masses, 
as labelled in the figure, and the points with triangles mark the
states where fagmentation conditions are met (see text).}
\label{fig:fesinT}\end{figure}
\begin{figure}
\center{\epsfig{figure=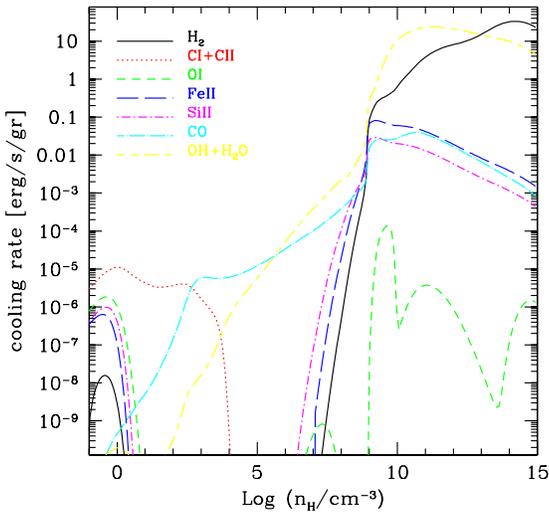, height=8.5cm}}
\caption{Individual metal fine-structure lines (FeII, SiII,
CII+CI, OI) and molecular lines (H$_2$, CO, OH+H$_2$O) contributions to the
cooling rate of a gas cloud with initial metallicity $Z=10^{-2} Z_{\odot}$ (see Fig.\ref{fig:fesinT}).}
\label{fig:fesicool}\end{figure}

\subsection{Dust cooling} 
\label{sec:dustcool}

Dust grains affect the thermal evolution of the collapsing gas cloud when the energy
transfer between gas and dust due to collisions is radiated away. Dust thermal emission rate can be expressed as,

\begin{equation}
\Lambda_{\rm gr} \mbox{[erg/s/gr]} = 4 \, \sigma  \, T_{\rm gr}^4  \,\kappa_{\rm P}   \,\beta_{\rm esc}  \, \frac{\rho_{\rm gr}}{\rho}, 
\label{eq:tgrain}
\end{equation}
\noindent
where $\sigma$ is the Stefan-Boltzmann constant, $T_{\rm gr}$ is the dust temperature,
$\kappa_{\rm P}$ is the Planck mean opacity of dust grains per
unit dust mass, $\beta_{\rm esc}$ is the photon escape probability, and the dust mass density, 
$\rho_{\rm gr}$, can be expressed as $\rho \, {\cal D}$, with ${\cal D}$ indicating the dust-to-gas ratio.
Following Omukai (2000), the photon escape probability can be written as,
\be
\beta_{\rm esc} = {\rm min}(1,\tau^{-2}) \qquad \mbox{with} \qquad \tau = \kappa_{\rm P} \rho \lambda_J,
\ee
\noindent
where the optical depth is estimated across one local Jeans length (roughly the
size of the central core region in the Penston-Larson similarity solution).
The Planck mean opacity per unit dust mass is,
\begin{equation}
{\kappa}_{\rm P} = \frac{ \pi}{\sigma T_{\rm gr}^4} \int_0^{\infty} B_{\nu}(T_{\rm gr}) \kappa_\nu d\nu
\label{eq:kplanck}
\end{equation}
\noindent
where $\kappa_\nu$ is the absorption coefficient for the frequency $\nu$
per unit dust mass and $B_{\nu}(T_{\rm gr})$ is the Planck brightness (see Schneider et al. 2006).

The dust temperature, $T_{\rm gr}$, is determined by the balance between 
dust thermal emission and the dust heating rate due to
collisions with gas particles (Hollenbach \& McKee 1979), $H_{\rm gr} = \rho \Lambda_{\rm gr}$, with 

\begin{equation}
H_{\rm gr} \mbox{[erg/s/cm$^3$]}= \frac{n_{\rm gr}(2 \, k  \,T - 2  \,k  \,T_{\rm gr})}{t_{\rm coll}},
\label{eq:heat}
\end{equation}
\noindent
where $t_{\rm coll}^{-1} = n_{\rm H}\,\sigma_{\rm gr} \,\bar{v}_{\rm H}\,f$ is the average time between two successive
collisions, $n_{\rm gr}$ and $\sigma_{\rm gr}$ are the grain number density and cross section, $\bar{v}_{\rm H}$ is
the average speed of hydrogen nuclei and $f = \bar{n v}/n_{\rm H}\bar{v}_{\rm H} \approx 0.395$ measures the contribution of
other species\footnote{Since we are interested in very low gas metallicities ($10^{-7} Z_{\odot} \le Z \le 10^{-4} Z_{\odot}$)
we can safely neglect the contribution of heavy elements and write $f = 0.3536 + 0.5 y_{\rm He}$ with $y_{\rm He} = 
n_{\rm He}/n_{\rm H} = 0.083$, which corresponds to a He mass fraction of Y=0.25.}.

In the density range where collisional dust heating is relevant
($n_{\rm H} > 10^{12-13} \mbox{cm}^{-3}$), it is a good approximation to 
assume that the gas is fully molecular and follows a Maxwellian distribution
so that, 
\be
\bar{v}_{\rm H} = \left(\frac{8 \,k \,T}{\pi \,m_{\rm H}}\right)^{1/2}.
\label{eq:vel}
\ee
\noindent
If we further write,
\begin{equation} 
n_{\rm gr} \,\sigma_{\rm gr} = n_{\rm H} \,  m_{\rm H} (1 + 4 y_{\rm He}) \, S \, {\cal D} 
\label{eq:cross}
\end{equation}
\noindent
in eq.~(\ref{eq:heat}), where $S$ is the total grain cross section per unit mass of dust, 
the equation $\rho \Lambda_{\rm gr} = H_{\rm gr}$ appears
to be independent of the dust-to-gas ratio. Thus, $T_{\rm gr}$ depends only on gas temperature
and density and on grain specific properties 
(cross section and mean opacity, see Schneider et al. 2006 for further details).

\begin{figure*}
\center{\epsfig{figure=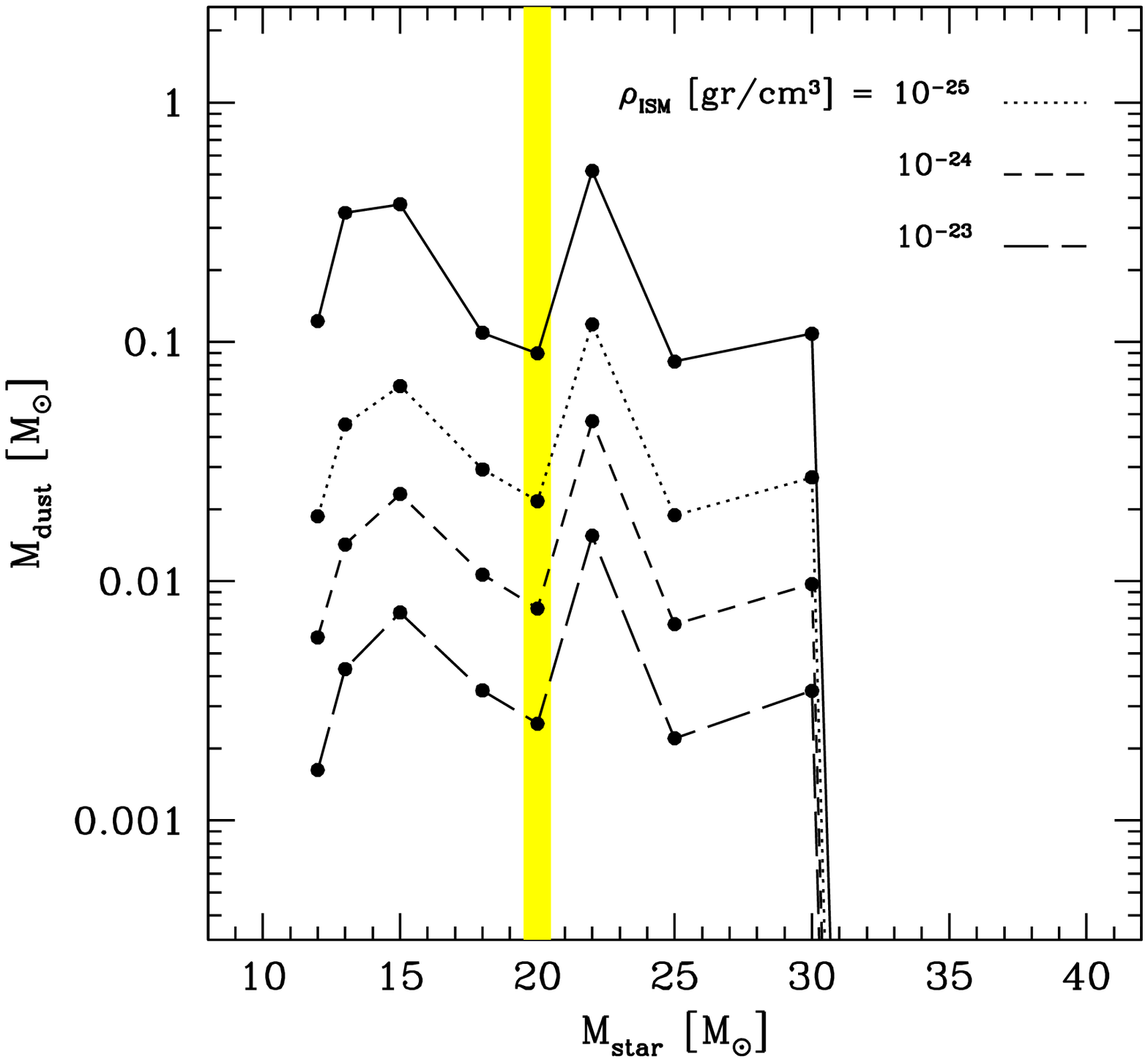, height=8.5cm}
\epsfig{figure=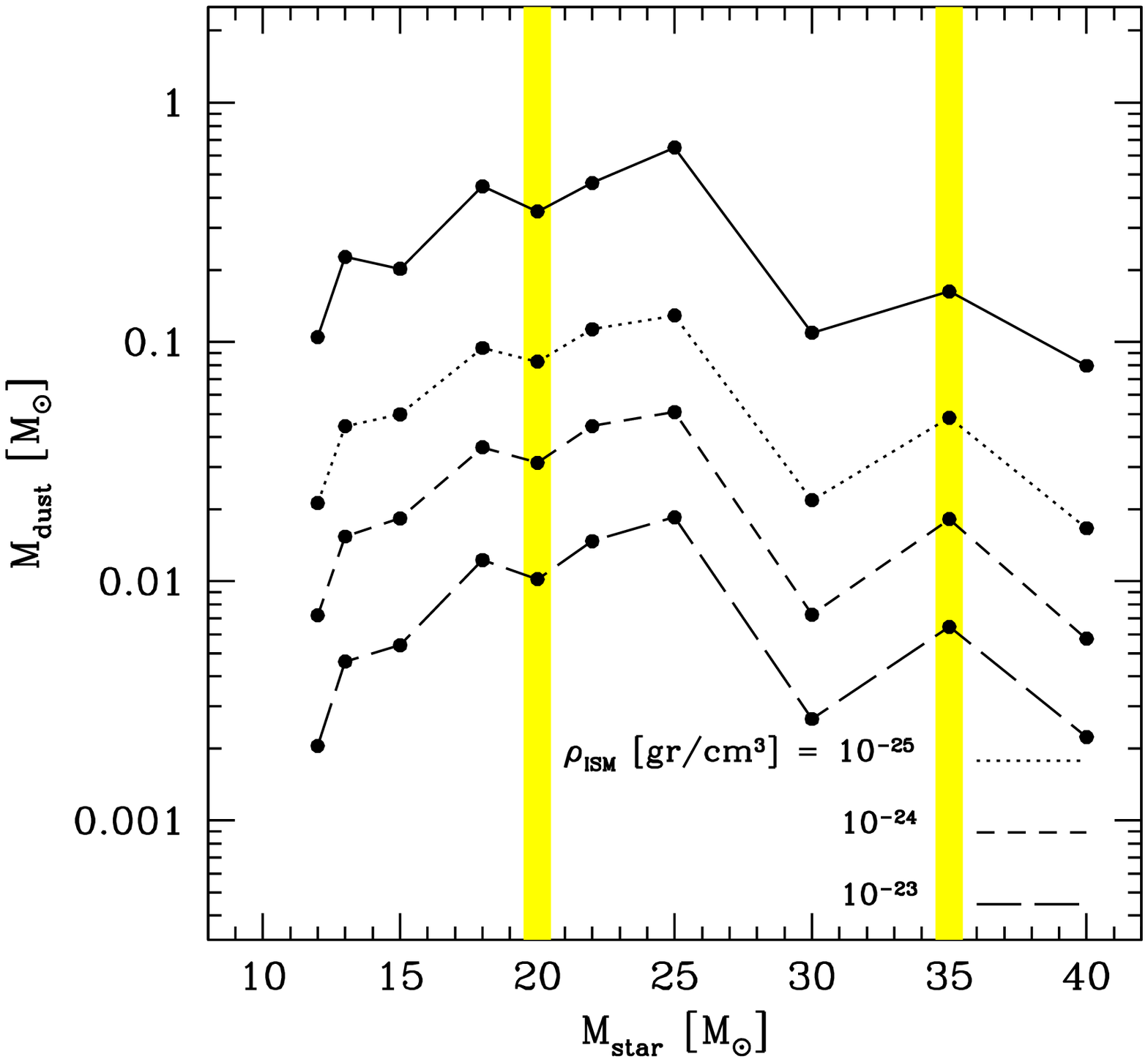, height=8.5cm}}
\caption{Mass of dust predicted by SN as a function of the progenitor stellar
mass for two initial metallicities: $Z=0$ (left panel) and
$Z=10^{-4} Z_{\odot}$ (right panel). In each panel, the solid line refers 
to the mass of dust predicted at the end of the condensation phase ({\bf norev}) 
and for the {\bf rev1}, {\bf rev2}, {\bf rev3} cases (dotted, dashed, long-dashed lines, respectively)
correponding to ISM densities of $10^{-25}$, $10^{-24}$ and $10^{-23}$~gr/cm$^{-3}$, respectively.}
\label{fig:snZU}
\end{figure*}

\begin{table*}
\begin{center}
\caption{Ejecta composition of the three SN models; for each of these we show the depletion factor, 
$f_{\rm dep}$ (first column), the mass of dust grains, $M_{\rm dust}$ (second column), 
the total mass of gas-phase metals, $M_{\rm met}$ (third column) and the mass of O and C atoms 
(fourth and fifth columns, respectively) in the {\bf norev}, {\bf rev1}, {\bf rev2}, and {\bf rev3} 
cases. All masses are in solar units. The last column shows the total dust cross section per unit dust mass
(see section \ref{sec:dustcool}).}
\label{table:ejectacomp}
\begin{tabular}{l|l|l|l|l|l|l|l}\hline
SN progenitor  & models  & $f_{\rm dep}$ &  $M_{\rm dust}$ & $M_{\rm met}$ & $M_{\rm O}$ & $M_{\rm C}$ & S [10$^5$ cm$^2$/gr] \\ \hline
$Z=0$, 20 $M_{\odot}$ & {\bf norev} & $0.91$ & $8.96 \times 10^{-2}$ & $8.93 \times 10^{-3}$ & $8.92 \times 10^{-3}$ & 0 & 4.71 \\
                      & {\bf rev1}  & $0.22$ & $2.16 \times 10^{-2}$ & $7.69 \times 10^{-2}$ & $8.92 \times 10^{-3}$ & $6.8 \times 10^{-2}$ & 5.24\\
                      & {\bf rev2}  & $7.81 \times 10^{-2}$ & $7.69 \times 10^{-3}$ & $9.08 \times 10^{-2}$ & $8.92 \times 10^{-3}$ & $8.19 \times 10^{-2}$ & 5.37 \\
                      & {\bf rev3}  & $2.58 \times 10^{-2}$ & $2.54 \times 10^{-3}$ & $9.60 \times 10^{-2}$ & $8.92 \times 10^{-3}$ & $8.71 \times 10^{-2}$ & 5.21 \\
\hline
$Z=10^{-4} Z_{\odot}$, 20 $M_{\odot}$ 	& {\bf norev} & 0.17                & 0.35 & $1.72$  & $1.47$ & 0 & 2.22 \\
                      			& {\bf rev1} &  $3.99 \times 10^{-2}$ & $8.27 \times 10^{-2}$ & $1.99$ & $1.50$  & $0.14$ & 2.44 \\
                      			& {\bf rev2} &  $1.51 \times 10^{-2}$ & $3.13 \times 10^{-2}$ & $2.04$ & $1.50$  & $0.19$ & 2.50 \\ 
                      			& {\bf rev3} &  $4.92 \times 10^{-3}$ & $1.02 \times 10^{-2}$ & $2.06$ & $1.50$  & $0.21$ & 2.50 \\
\hline
$Z=10^{-4} Z_{\odot}$, 35 $M_{\odot}$   & {\bf norev} & $0.87$                & $0.16$                & $2.45 \times 10^{-2}$ & $2.45 \times 10^{-2}$ & 0 & 2.35 \\
                      			& {\bf rev1}  & $0.26$                & $4.82 \times 10^{-2}$ & $0.14$ & $2.45 \times 10^{-2}$ & $0.11$ & 2.62 \\
                      			& {\bf rev2}  & $9.70 \times 10^{-2}$ & $1.82 \times 10^{-2}$ & $0.17$ & $2.45 \times 10^{-2}$ & $0.14$ & 2.68 \\
                      			& {\bf rev3}  & $3.44 \times 10^{-2}$ & $6.44 \times 10^{-3}$ & $0.18$ & $2.45 \times 10^{-2}$ & $0.16$ & 2.62 \\
\hline
\end{tabular}
\end{center}
\end{table*}

\section{Fragmentation by metal--cooling: effects of Fe and Si lines}
\label{sec:fesicool}

In this section, we discuss the thermal evolution of collapsing gas clouds when
only molecules and metals contribute to the cooling rate in eq.~(\ref{eq:cool}).
Thus, we assume that no dust is present. For simplicity, we adopt an ISM with 
the solar elemental composition (Anders \& Grevesse 1989) and the mass 
of each element is decreased proportionally to the total metallicity. 

To estimate the effects of FeII and SiII lines on gas cooling,
we do not solve the full Fe and Si chemistry (as we do for the other elements)
but we simply assume that all the Fe and Si contributes to cooling.
Fig.~\ref{fig:fesinT} shows the thermal evolution of collapsing gas clouds
with varying initial metallicities ($Z=0, 10^{-4} Z_{\odot}, 10^{-3} Z_{\odot}, 10^{-2} Z_{\odot}$,
from top to bottom). 
For the $Z > 0$ models, thin lines represent the predictions obtained with Omukai et al. (2005)
models and thick lines show the current models with the inclusion of SiII and FeII line--cooling.
As it is apparent from the figure, the difference is very small and confined to the density range
$n_{\rm H} < 10^6$~cm$^{-3}$. In fact, even for the highest metallicity model that we have considered, $Z=10^{-2} Z_{\odot}$, 
the cooling rate
is dominated by molecular lines, specifically by H$_2$O and OH. 
This is shown in Fig.~\ref{fig:fesicool}, 
where we compare the individual contributions to the cooling rate.
Although FeII and SiII are the most important coolants among
metal fine--structure lines in some density ranges, their 
contribution to the total cooling rate is
smaller than that provided by molecules.

\begin{figure*}
\center{\epsfig{figure=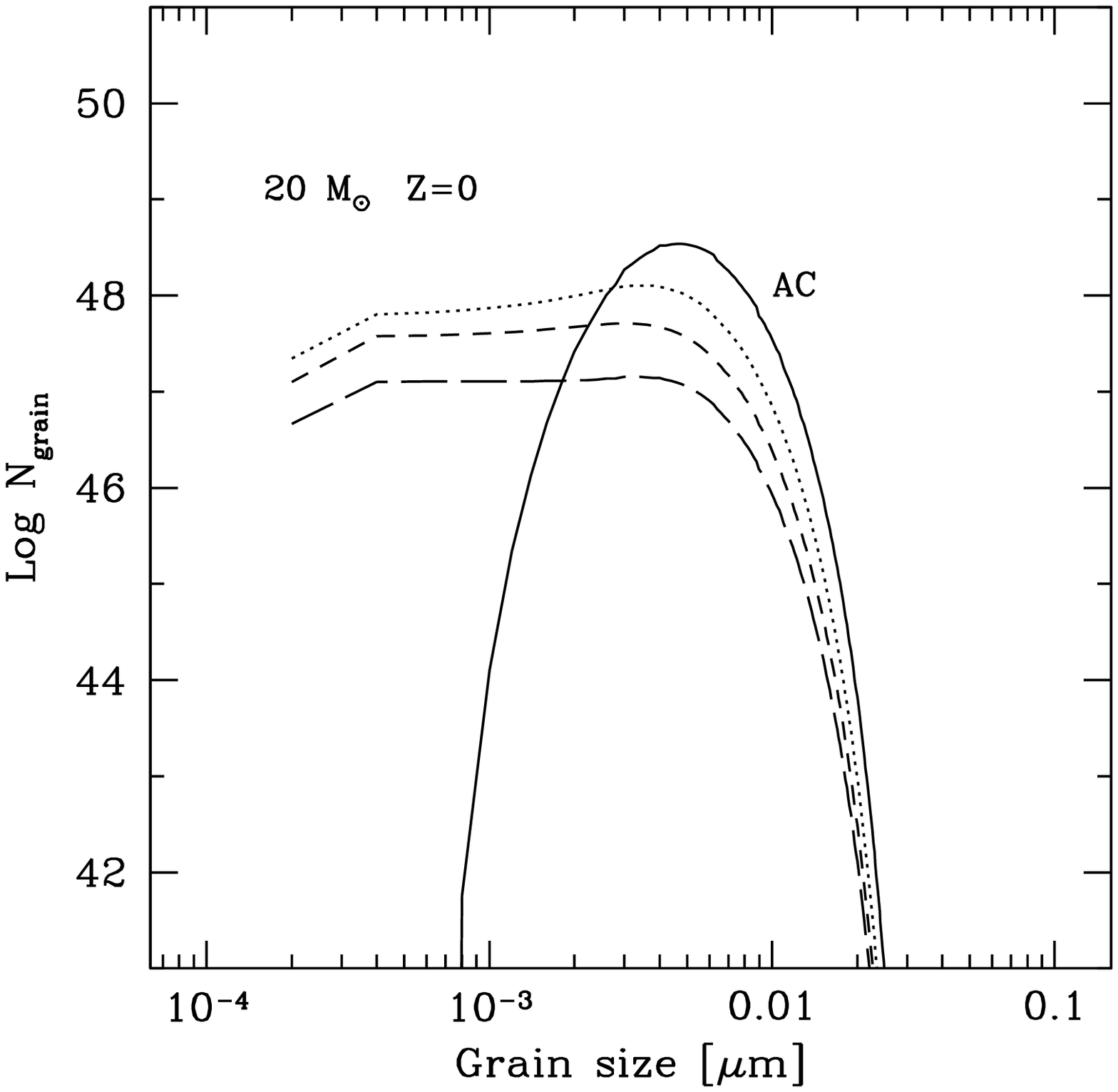, height=5.8cm}
\epsfig{figure=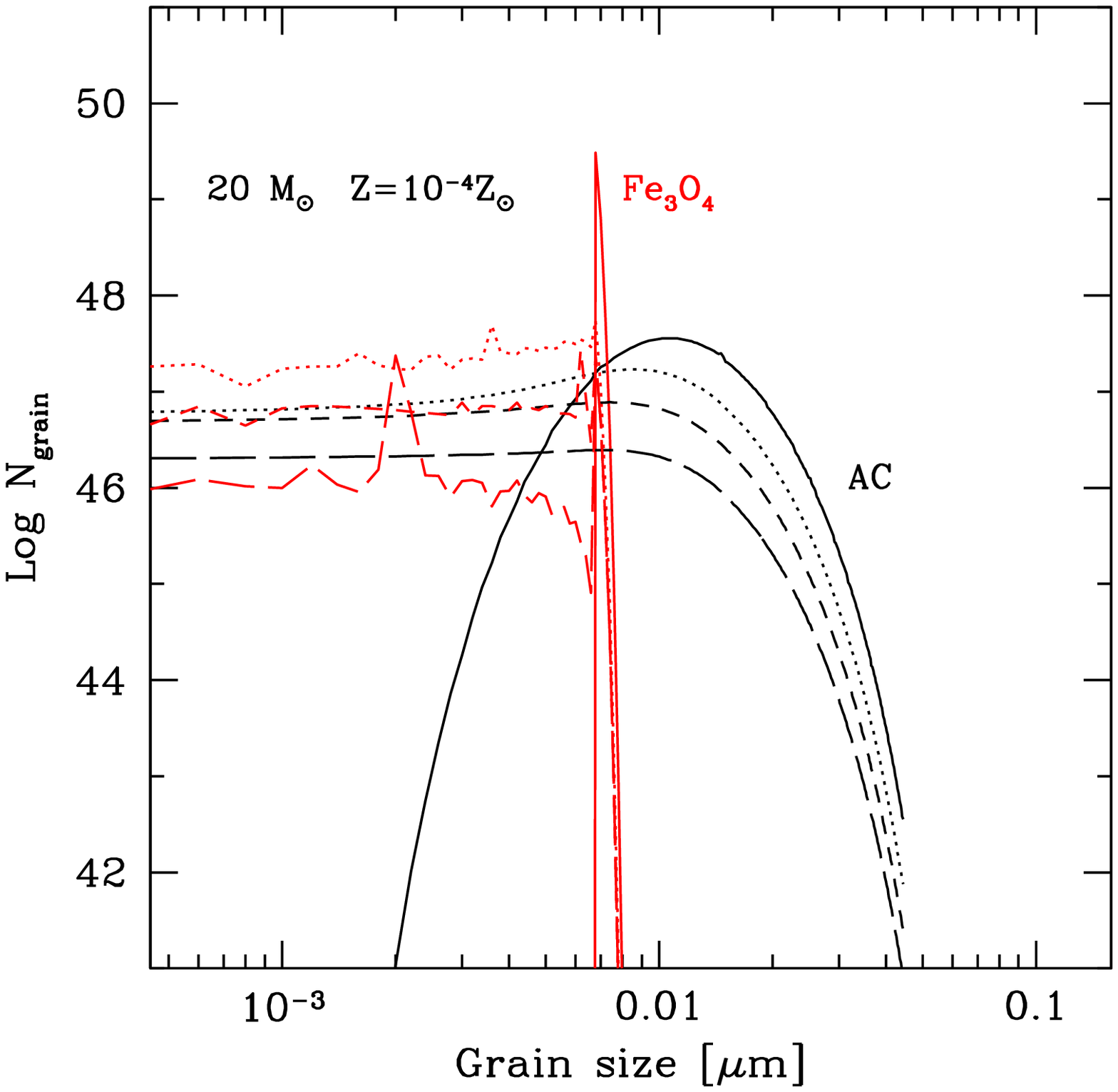, height=5.8cm}
\epsfig{figure=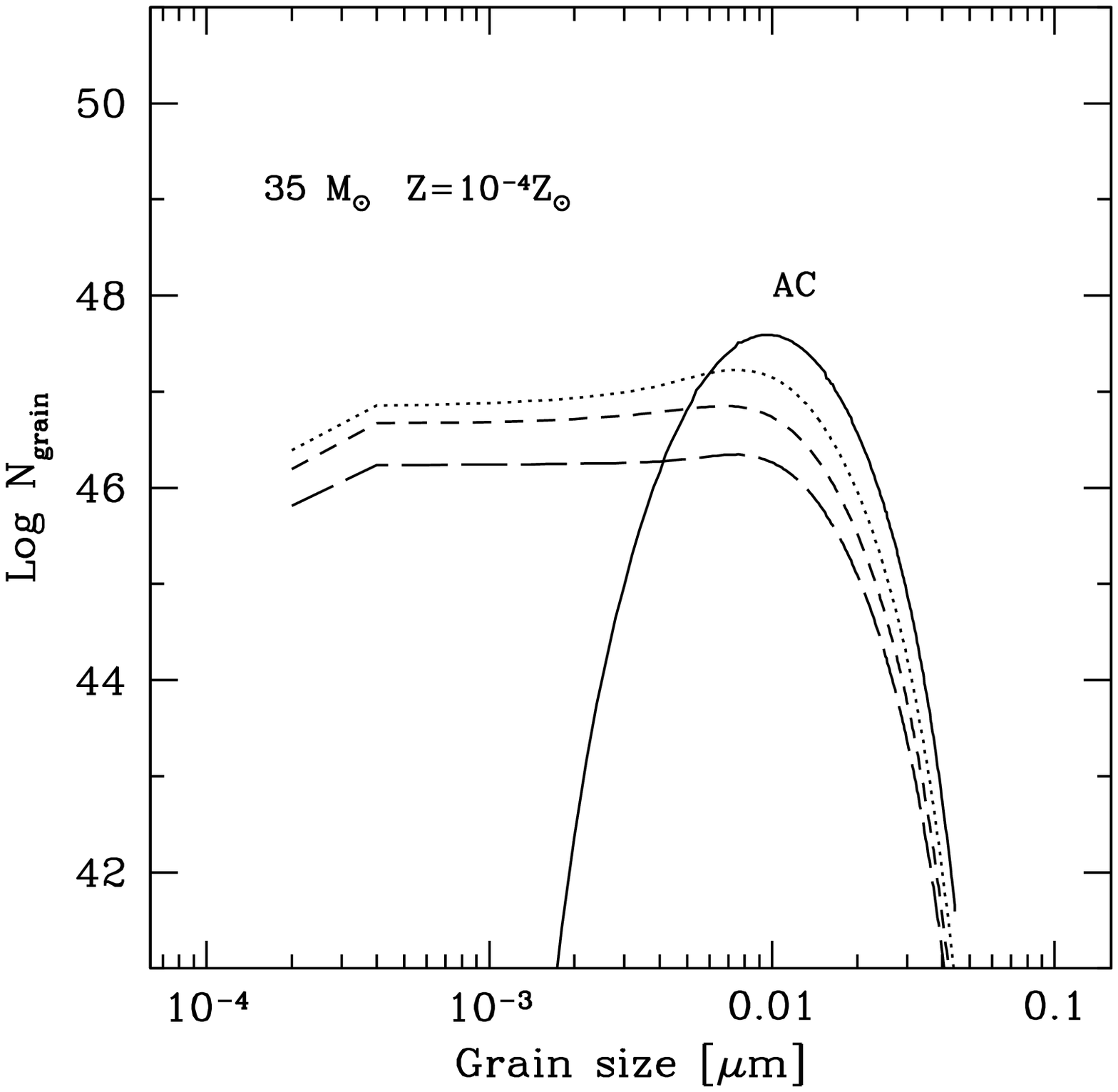, height=5.8cm}}
\caption{Size distribution function of the dominant dust species resulting from the 
explosion of a $Z=0$, $M_{\rm star} = 20 M_{\odot}$ (left panel),
$Z=10^{-4}Z_{\odot}$, $M_{\rm star} = 20 M_{\odot}$ (central panel), and
$Z=10^{-4}Z_{\odot}$, $M_{\rm star} = 35 M_{\odot}$ (right panel). Solid lines
show the distribution at the end of the condensation phase ({\bf norev}) and
dotted, dashed, long-dashed lines illustrate the effect of reverse shocks with 
increasing strength ({\bf rev1}, {\bf rev2}, {\bf rev3}, respectively).}
\label{fig:sizecomp}
\end{figure*}

In Fig.\ref{fig:fesinT}, the diagonal dotted lines represent constant Jeans mass values in the
$(n_{\rm H}, T)$ plane. These are computed using the relation,
\begin{eqnarray}
&&M_{\rm J} =  \rho \lambda_{\rm J}^3 \\ \nonumber
& = & 1.69 \times 10^3 M_\odot \left(\frac{\mu}{1.23}\right)^{-3/2} \left(\frac{T}{200\mbox{K}}\right)^{3/2} 
\left(\frac{n_{\rm H}}{10^4 \mbox{cm}^3} \right)^{-1/2},
\label{eq:mjeans}
\end{eqnarray}
where the Jeans length is,
\begin{equation} 
\lambda_{\rm J} = \left(\frac{\pi k_{\rm B} T}{G \mu m_{\rm H} \rho}\right)^{1/2}.
\end{equation}
\noindent
The points with triangles mark the states where fragmentation conditions are met. 
Following Schneider \& Omukai (2010), we identify the fragmentation epochs by 
requiring that the adiabatic index becomes $\gamma > 0.97$ after a phase of cooling
(where $\gamma < 1$). To eliminate ``false" fragmentation points,
where $\gamma$ is less than 1 only for a short period of time, we impose
that $\gamma < 0.8$ during the cooling phase preceding the
fragmentation epoch (see Schneider \& Omukai 2010 for more details). 

It is clear from the figure that the minimum fragmentation masses that can
be achieved by metal and molecular line--cooling are $\ge 10 M_{\odot}$, even
for gas clouds with initial metallicity as high as $Z = 10^{-2} Z_{\odot}$. Thus, the
additional contribution of Fe and Si lines to the chemical network and cooling
rate do not change the main conclusions of previous investigations, based
on the model presented in Omukai et al. (2005). 
In particular, the formation of solar or sub--solar mass fragments
requires the presence of dust grains and dust-induced cooling at densities
$n_{\rm H} > 10^{12}$cm$^{-3}$.\\

\section{Model grid of core-collapse SN}
\label{sec:sngrid}

Figure~\ref{fig:snZU} shows the masses of dust produced by core
collapse SN as a function of the progenitor star mass 
and initial metallicities equal to $Z=0$ (left panel) and 
$Z=10^{-4} Z_{\odot}$ (right panel). The solid lines show dust masses at
the end of the condensation phase (no reverse shock model) according to the model developed 
by Todini \& Ferrara (2001) and revisited by Bianchi \& Schneider (2007).
Here we refer to the fiducial case, where we assume that the first seed
critical dust clusters are formed with a minimum of two monomers and
that the sticking coefficient is equal to 1 (we refer to Bianchi \& Schneider
2007 for more details). We see that the most massive SN ($M_{\rm star} > 25 M_{\odot}$)
are less efficient sources of dust due to the strong fallback of metals during the
explosion. This is particularly true for the primordial case, where metal (and
therefore dust) production is zero beyond a progenitor stellar mass of $30 M_{\odot}$.
The dotted, dashed, and long-dashed lines show the masses of dust after the passage
of the reverse shock when the surrounding ISM density is assumed to be equal to
$10^{-25}$, $10^{-24}$ and $10^{-23}$~gr/cm$^{-3}$, respectively.
The reverse shock is modelled following
the prescription of Bianchi \& Schneider (2007) and dust destruction is caused by 
both thermal and non-thermal sputtering (see the original paper for more details). 
The fraction of the initial dust mass which is destroyed and returns back to the
gas phase increases with the ISM density. Hereafter, we will refer to the ejecta properties
at the end of the condensation phase as the {\bf norev} case; 
with {\bf rev1}, {\bf rev2}, {\bf rev3} cases we will indicate the models
with reverse shock of increasing strength (increasing ISM density). 

As it is reasonable to expect, the mass of dust released by SN increases with the
metallicity of the progenitor stars. We focus here on three representative SN models 
(highlighted in the figure): 
a $Z=0$, 20 $M_{\odot}$ progenitor ($20 M_{\odot}$, 0) and two $Z=10^{-4} Z_{\odot}$ progenitors with masses
of $20$ ($20 M_{\odot}$, $10^{-4} Z_{\odot}$) and $35 M_{\odot}$ ($35 M_{\odot}$, $10^{-4} Z_{\odot}$). 
Table~\ref{table:ejectacomp} describes the final ejecta
composition for these three SN in the {\bf norev}, {\bf rev1}, {\bf rev2} and {\bf rev3} cases.
The first column lists the depletion factor, $f_{\rm dep}$, with values  
in the range $4.92 \times 10^{-3} \le f_{\rm dep} \le 0.91$, decreasing with reverse shock strength. 
Thus, the conditions we will explore span the value appropriated for the local ISM ($f_{\rm dep}=0.49$) 
as well as values which are 2 dex smaller. Note that the depletion factors are systematically larger
for the ($20 M_{\odot}$, 0) and the ($35 M_{\odot}$, $10^{-4} Z_{\odot}$) SN progenitors, 
despite the smaller final dust and metal yields with respect to the SN progenitor ($20 M_{\odot}$, $10^{-4} Z_{\odot}$).

The elemental composition of the ejecta depends on the SN progenitor and reverse shock model.
As it can be seen from Table~\ref{table:ejectacomp}, oxygen is always the dominant element at the end of the condensation phase 
({\bf norev} cases) and carbon is returned to the gas phase after the partial destruction of graphite grains
by the reverse shock ({\bf rev1}-{\bf rev3} cases).
We therefore expect that the thermal properties of gas enriched by these 
ejecta will depend on $f_{\rm dep}$ due to variations both in the abundance of dust
grains and in the composition of gas-phase metals.  

\begin{figure}
\center{\epsfig{figure=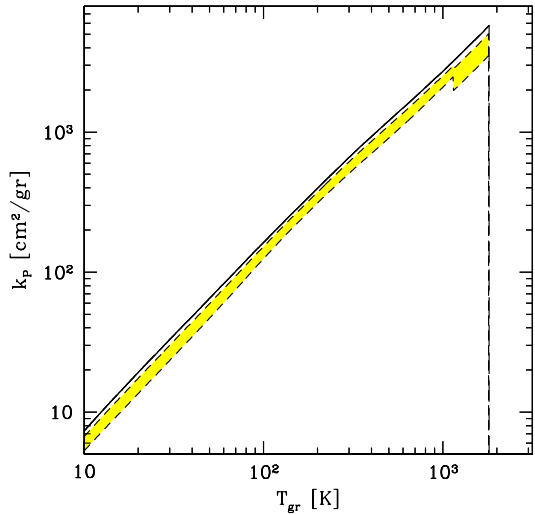, height=8.5cm}}
\caption{Planck mean opacity per unit dust mass as a function of  
grain temperature. The solid line refers to models $Z=0$, $M_{\rm star} = 20 M_{\odot}$,
and $Z=10^{-4}Z_{\odot}$, $M_{\rm star} = 35 M_{\odot}$ which have very
similar dust properties. Also, for these two models we only show the opacity at the end
of the condensation phase because the passage of the reverse shock has a negligible effect.
The dashed region illustrates the maximum ({\bf norev}) and minimum ({\bf rev3}) 
Planck mean opacity per unit mass for the $Z=10^{-4}Z_{\odot}$, $M_{\rm star} = 20 M_{\odot}$ model.
Line breaks are due to grain sublimation (see text).}
\label{fig:opticalprop}
\end{figure}

The reverse shock significantly affects the grain size distribution function,
decreasing the number of larger grains and flattening the distribution towards smaller grain sizes.
This is shown in Fig.\ref{fig:sizecomp} where, for each model, the dominant grain species have been reported. 
Only amorphous carbon (AC) grains form in the ejecta of the ($20 M_{\odot}$,0) and the 
($35 M_{\odot}$,$10^{-4} Z_{\odot}$) progenitors;
additional grain species form in the ejecta of the ($20 M_{\odot}$,$10^{-4} Z_{\odot}$) progenitor, 
among which Fe$_3$O$_4$, MgSiO$_3$,SiO$_2$, and Al$_2$O$_3$ (see also Bianchi \& Schneider 2007 for more details). 
Thus, the former two SN progenitors share very similar dust properties. This is reflected by their corresponding 
grain cross sections per unit dust mass (see the last column in Table~\ref{table:ejectacomp}) and 
Planck mean opacities (see eq.~\ref{eq:kplanck}) which are plotted in Fig.~\ref{fig:opticalprop} 
as a function of dust grain temperature. 
The solid lines show the predicted opacity for ($20 M_{\odot}$,0) and ($35 M_{\odot},10^{-4} Z_{\odot}$) 
progenitors with no variations induced by dust destruction in the reverse shock; the shaded region indicates the maximum ({\bf norev} case)
and minimum ({\bf rev3} case) Planck mean opacity predicted for the ($20 M_{\odot},10^{-4}Z_{\odot}$) progenitor. 
The line breaks are due to grain sublimation, which occurs at a temperature of $1810$~K  for AC grains and of $1140$~K for the
additional grain species (Kozasa, Hasegawa, \& Nomoto 1989).  
Since the grain cross section and Planck mean opacity control the efficiency of dust cooling (see Section \ref{sec:dustcool}), 
in what follows we restrict the discussion to the most dissimilar SN progenitors, hence a $20 M_{\odot}$ star with metallicities $Z=0$ and
$Z=10^{-4} Z_{\odot}$.

\section{Results}
\label{sec:results}
\begin{figure*}
\center{\epsfig{figure=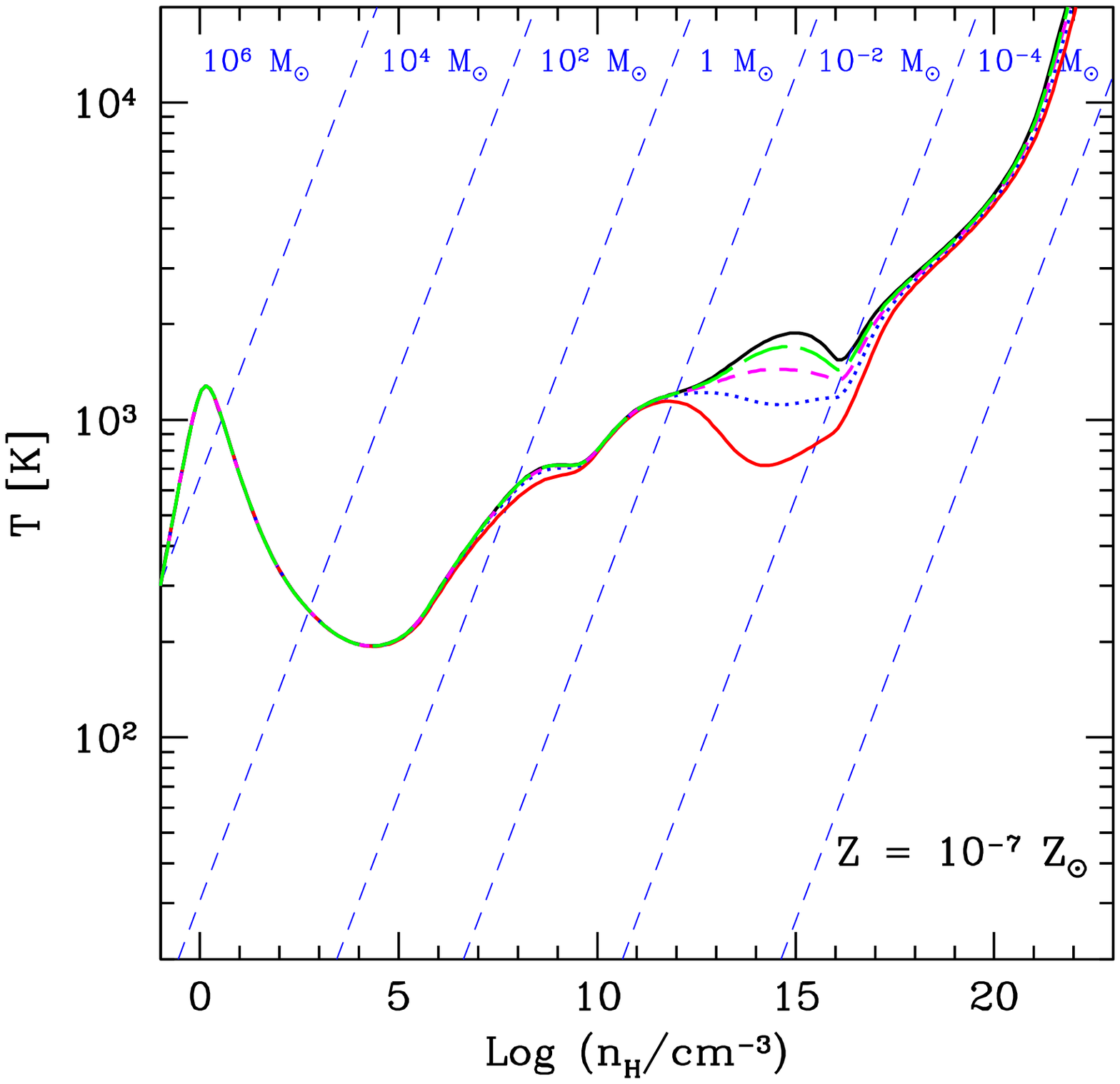, height=8.5cm}
\epsfig{figure=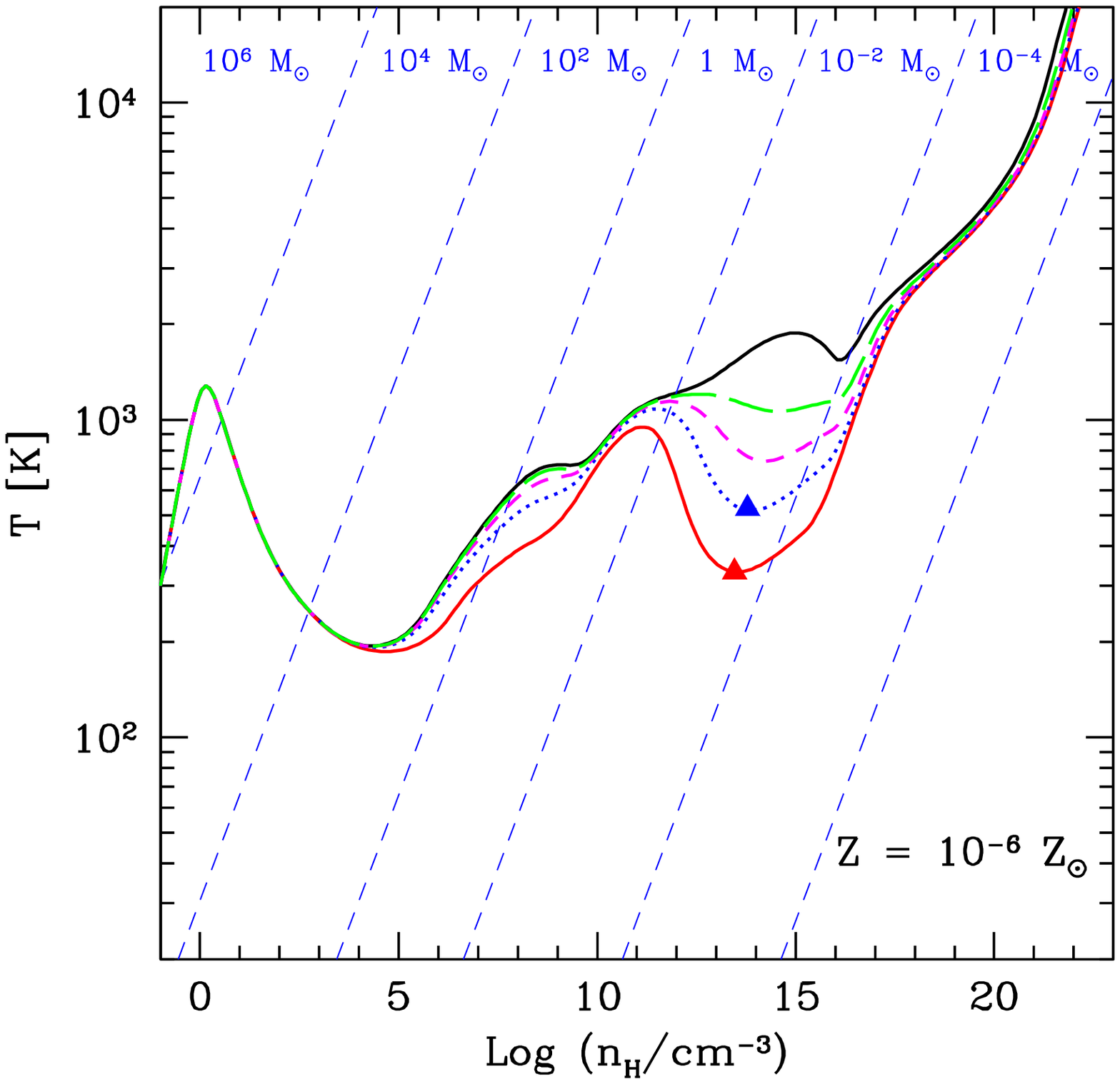, height=8.5cm}
\epsfig{figure=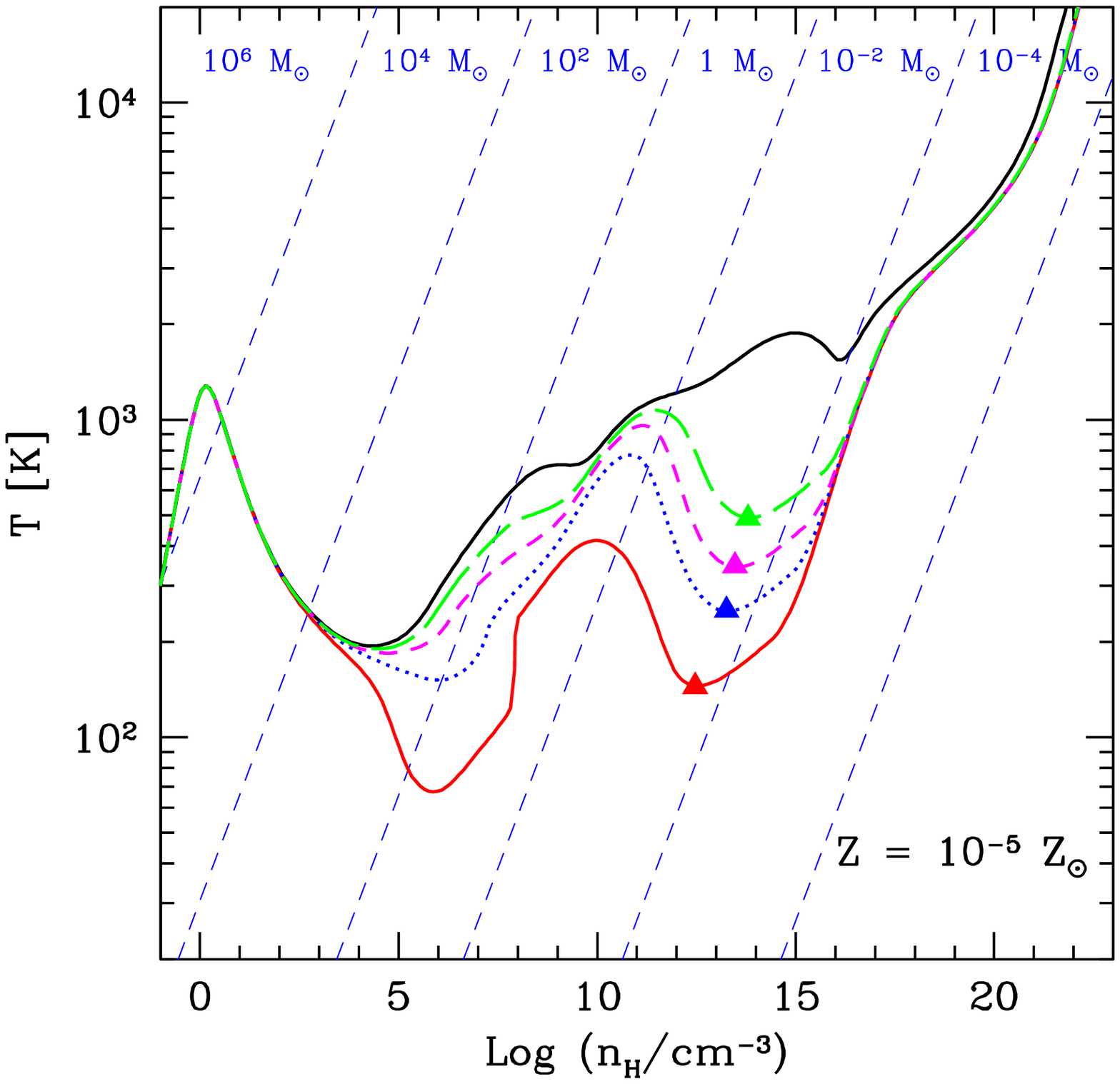, height=8.5cm}
\epsfig{figure=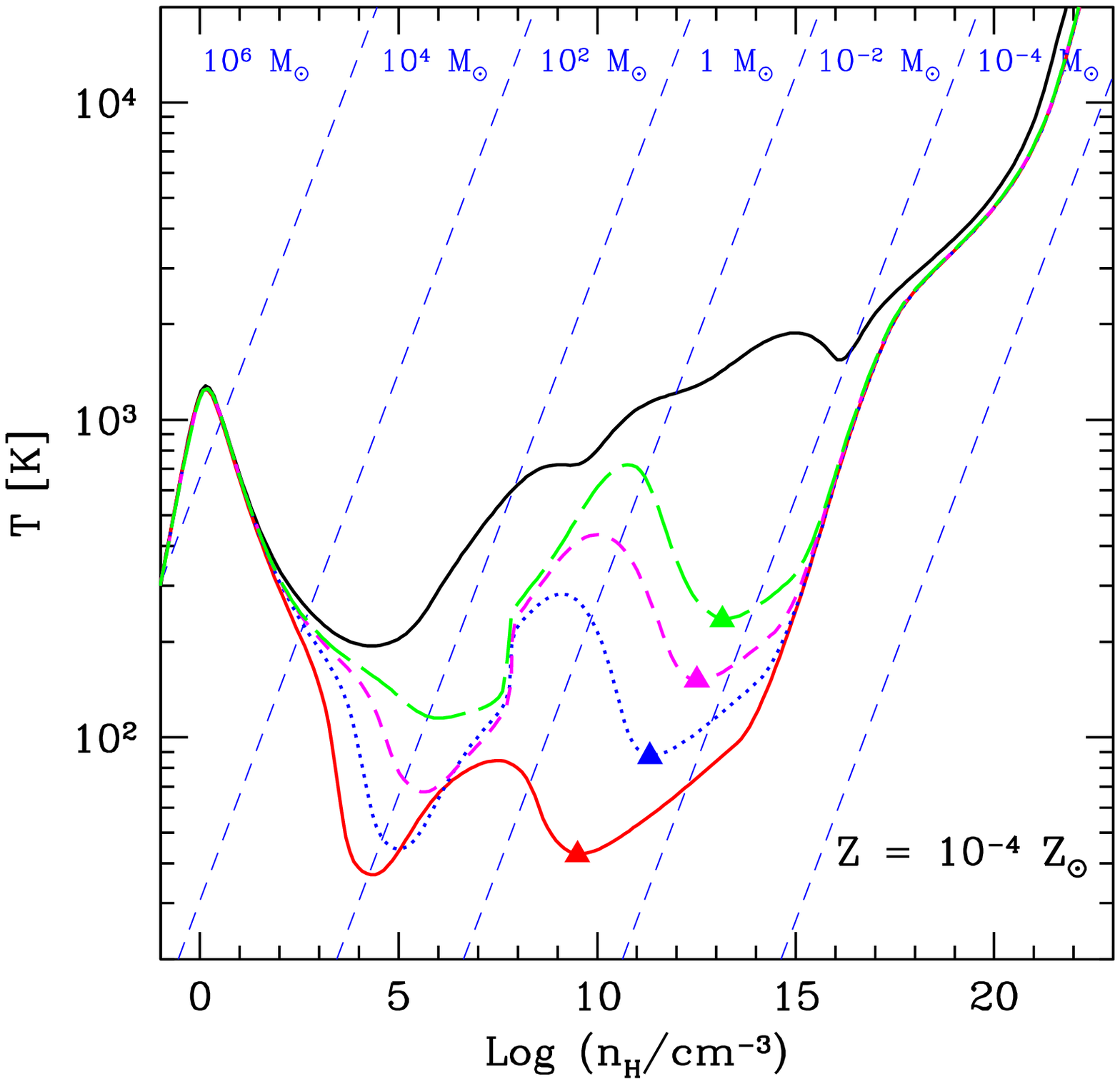, height=8.5cm}}
\caption{Thermal evolution of clouds enriched by the ($20 M_\odot$,0) SN model.
In each panel, we have assumed that the gas has been pre-enriched to a total metallicity
of $10^{-7} Z_{\odot}, 10^{-6} Z_{\odot}, 10^{-5} Z_{\odot}$,
and $10^{-4} Z_{\odot}$ (from top left to bottom right). 
In each panel, we show the evolution for four values of $f_{\rm dep}$ 
corresponding to the {\bf norev, rev1, rev2}, and {\bf rev3} cases 
(bottom solid, dotted, short-dashed, long-dashed, respectively). 
For each of these cases, the ejecta composition is varied consistently, as
presented in Table~\ref{table:ejectacomp} and Figs.~\ref{fig:sizecomp}, \ref{fig:opticalprop}. 
As a reference model, we also show the evolution
predicted for a purely primordial gas cloud (top solid line). The diagonal dashed lines represent 
lines of constant Jeans mass and the triangles identify the points where 
fragmentation conditions are met (see text).}
\label{fig:thermal20Z0}
\end{figure*}

\begin{figure*}
\center{\epsfig{figure=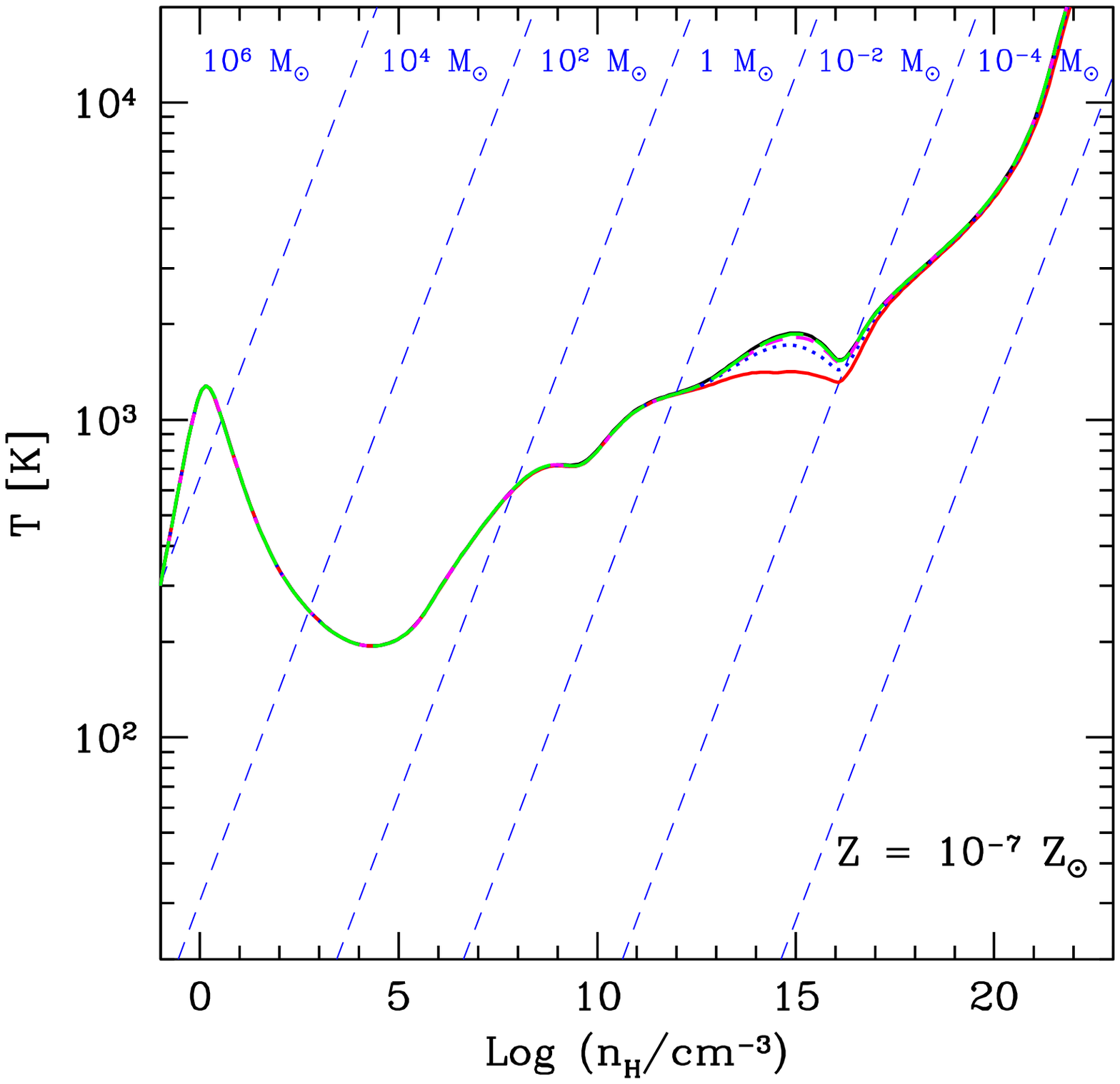, height=8.5cm}
\epsfig{figure=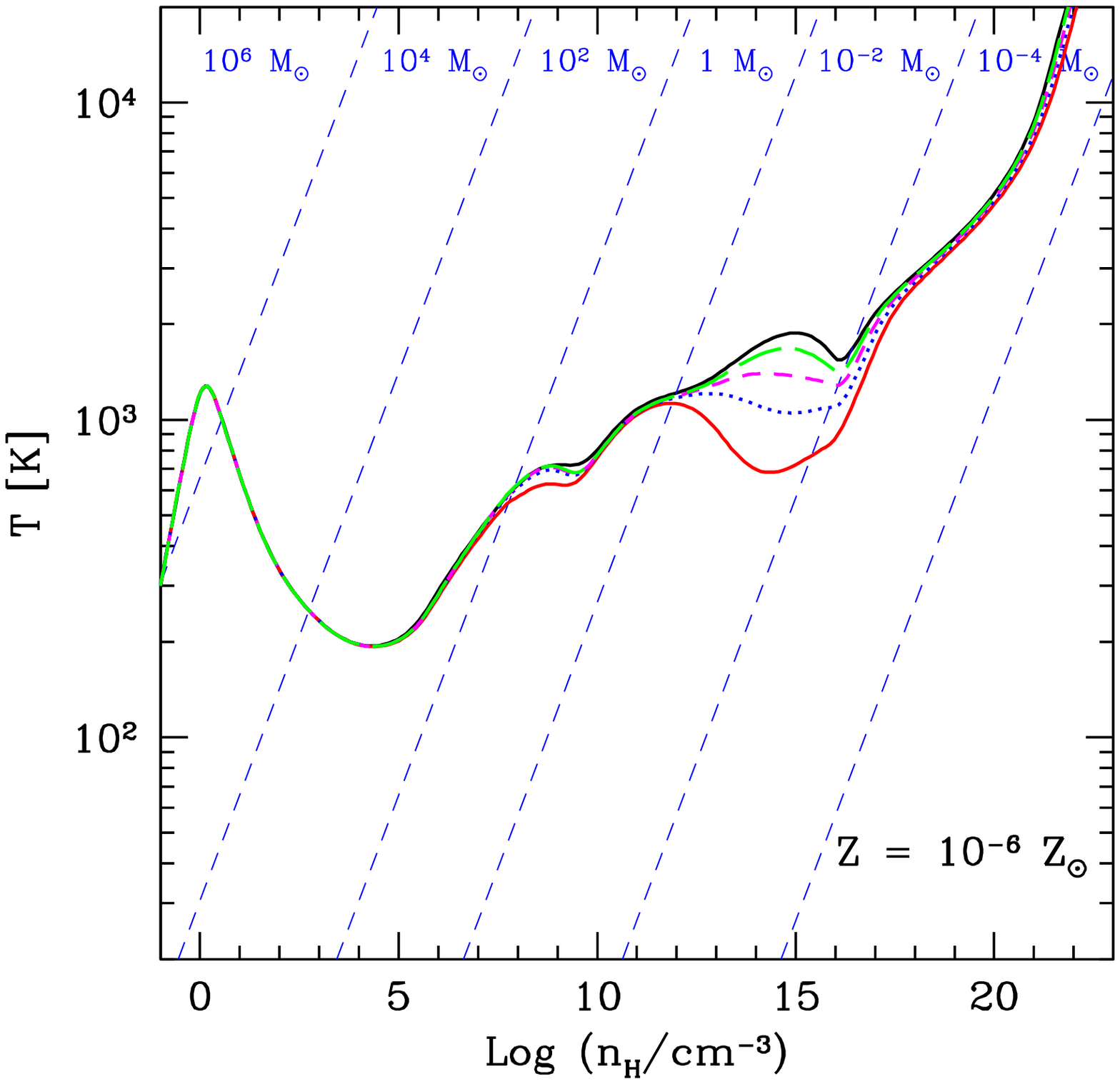, height=8.5cm}
\epsfig{figure=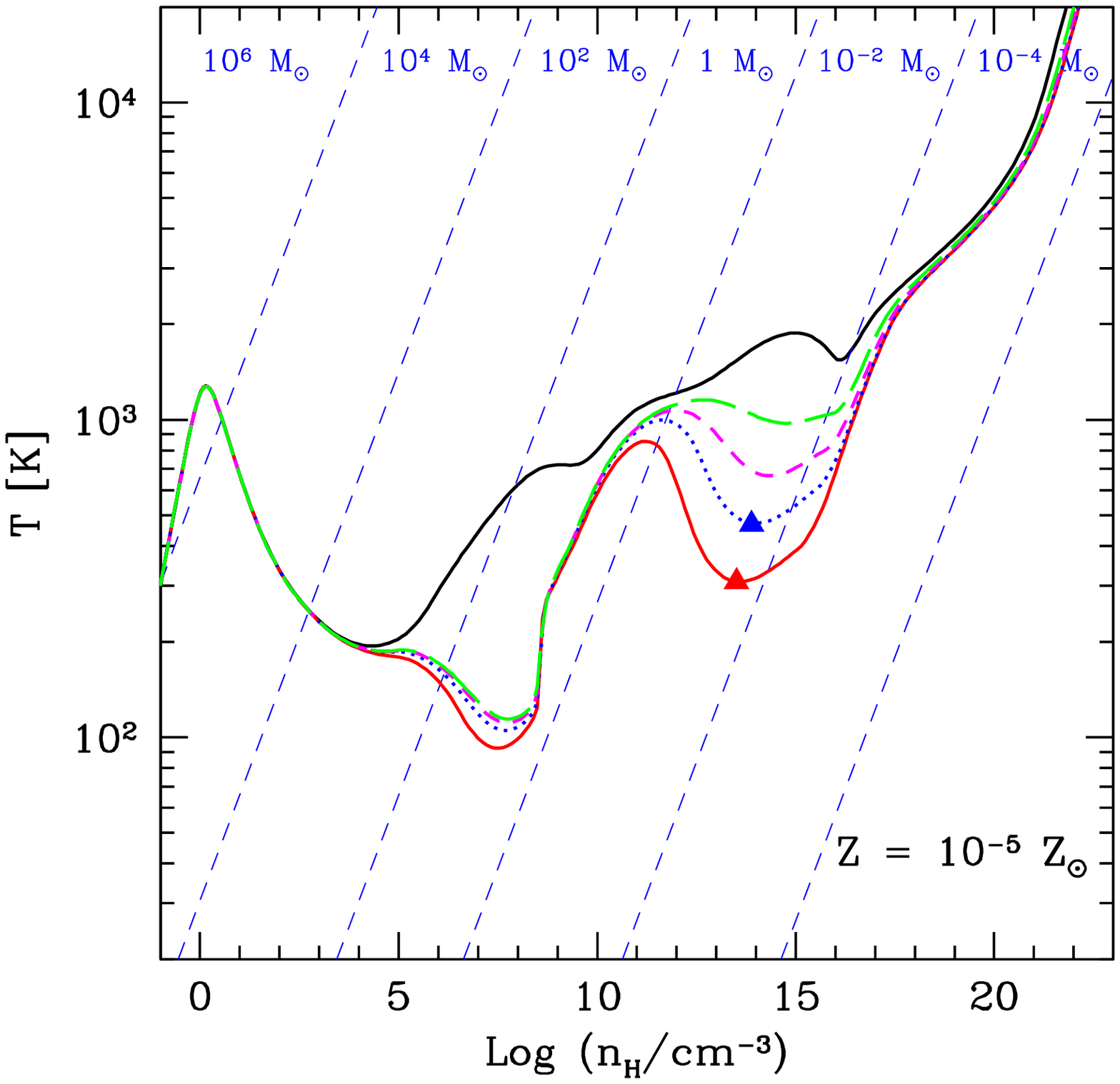, height=8.5cm}
\epsfig{figure=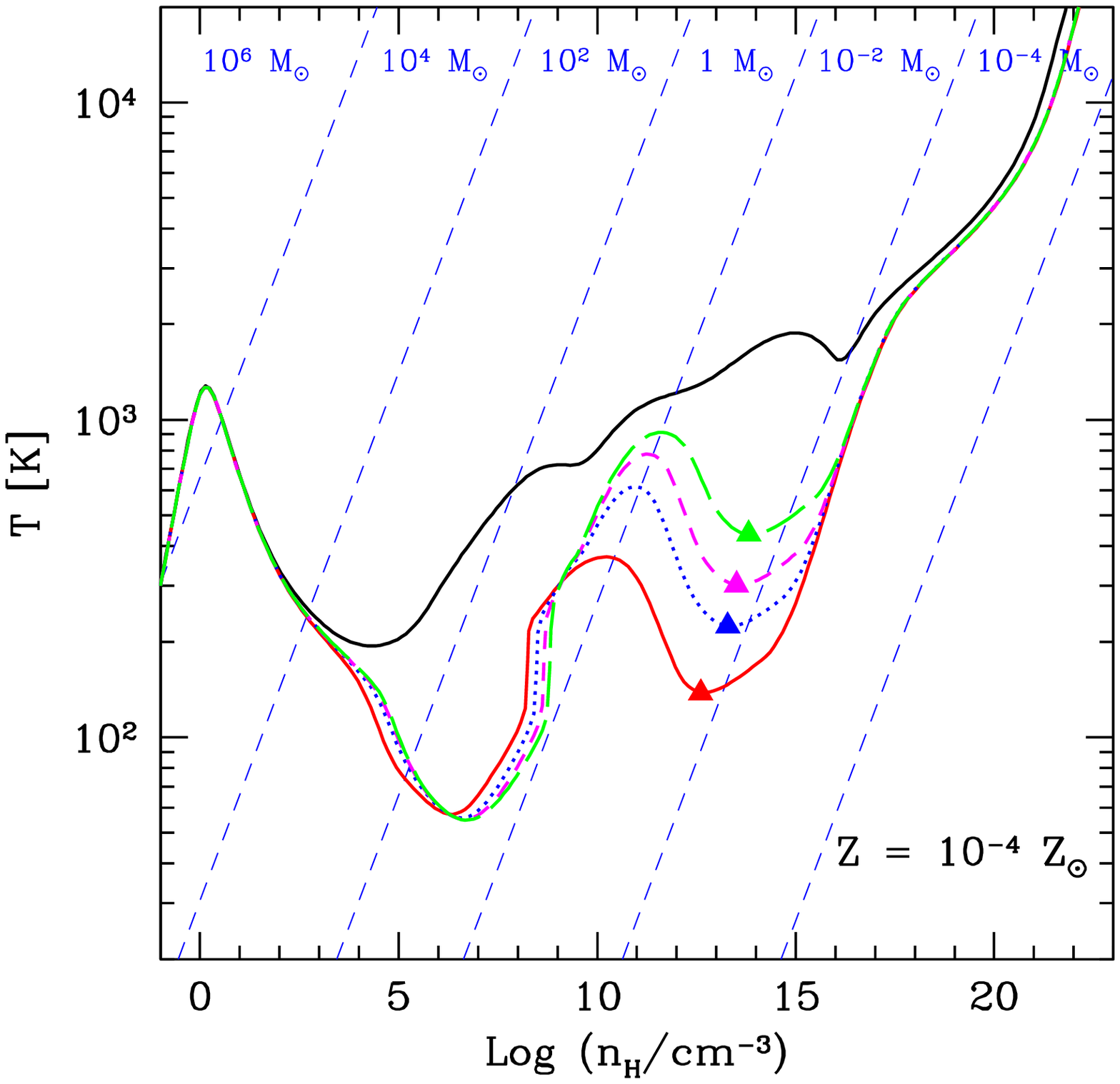, height=8.5cm}}
\caption{Same as Fig.~\ref{fig:thermal20Z0} but for the 
($20 M_{\odot},10^{-4}Z_\odot$) SN model.}
\label{fig:thermal20Z-4}
\end{figure*}

In this section we investigate the thermal evolution and fragmentation
of collapsing gas clouds enriched by each of the two representative supernova progenitors.
We assume that their ejecta uniformly enrich nearby collapsing gas clouds. Depending on
its total gas mass, the collapsing cloud can be enriched to different total metallicities, where
by total metallicity we mean the total mass of metals (in the gas-phase and in dust grains) 
relative to the mass of gas. 
Figs.\ref{fig:thermal20Z0} and \ref{fig:thermal20Z-4} show the thermal evolution of a collapsing
gas cloud with initial metallicity equal to  
$10^{-7} Z_{\odot}, 10^{-6} Z_{\odot}, 10^{-5} Z_{\odot}$, and $10^{-4} Z_{\odot}$ (from top 
left to bottom right); in each panel, we plot different reverse shock models varying the 
depletion factor and the ejecta composition self-consistently (from bottom to top {\bf norev, rev1, rev2, rev3} 
correspond to solid, dotted, short-dashed and long-dashed lines);  
as a reference case, the evolution predicted for a $Z=0$ gas is also shown (top solid line).  

The comparison among different panels shows that the deviation from the primordial gas 
evolution increases with metallicity; when $Z=10^{-7} Z_{\odot}$, 
this is limited to relatively large densities, $n_{\rm H} \ge 10^{12}$~cm$^{-3}$, but 
progressively extends to lower densities as the metallicity increases. 
At metallicities $10^{-7} Z_{\odot} \le Z \le 10^{-4} Z_{\odot}$ the evolution is 
dominated by dust through both direct cooling 
(high densities; $\ga 10^{11}{\rm cm^{-3}}$) and increased H$_2$ 
cooling due to H$_2$ formation on dust grains (lower densities; $\la 10^{9}{\rm cm^{-3}}$), 
as already discussed in Omukai et al. (2005) and Schneider et al. (2006). 
At each given metallicity the largest deviation from the primordial case is 
associated to the {\bf norev} case, characterized by the largest $f_{\rm dep}$, and
the evolution progressively tend to the primordial limit when lower depletion factors are considered
(models {\bf rev1, rev2} and {\bf rev3}). 

The points with triangles in Figs.\ref{fig:thermal20Z0} and 
\ref{fig:thermal20Z-4} mark the states where fragmentation conditions are met.
The lowest metallicity at which fragmentation occurs is 
$Z=10^{-6} Z_{\odot}$ for the ($20 M_{\odot},0$) SN
progenitor and it is 1 dex larger, $Z=10^{-5} Z_{\odot}$, for the 
($20 M_{\odot},10^{-4} Z_{\odot}$) SN progenitor.
In both cases, however, cloud fragmentation 
depends on the depletion factor and it is suppressed when the strength of the reverse 
shock leads to a too large destruction of dust grains ({\bf rev2} and {\bf rev3} cases). 

These features are all consistent with the existence of a minimum dust-to-gas
ratio above which fragmentation is activated. Given our definitions 
of metallicity, $Z$, and dust depletion factor, $f_{\rm dep}$, the dust-to-gas ratio 
can be written as ${\cal D} = Z f_{\rm dep}$. Thus, the minimum ${\cal D}$ is reached at
a lower total metallicity for models with larger depletion factors, such as the
($20 M_{\odot},0$) SN progenitor (see Table~\ref{table:ejectacomp}).

\begin{figure}
\center{\epsfig{figure=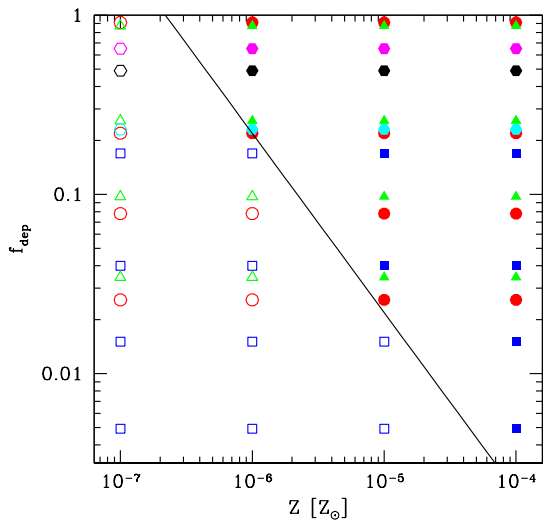, height=8.5cm}}
\caption{Fragmentation (filled symbols) and no-fragmentation (empty symbols)
conditions for each considered SN progenitors and reverse shock model. Red circles,
blue squares, and green triangles refer to ($20 M_\odot,0$), ($20 M_{\odot}, 10^{-4} Z_{\odot}$),
and ($35 M_{\odot}, 10^{-4} Z_{\odot}$) models, respectively. In addition, we also show the
SN and PISN models analyzed in Schneider et al. (2006, cyan and magenta exagons) and the
model which adopts local ISM conditions as in Omukai et al. (2005, black exagon).
The black solid line shows the predicted behaviour for a constant dust-to-gas ratio of
${\cal D}_{\rm cr} = 4.4 \times 10^{-9}$ (see text).}
\label{fig:dustcrit}
\end{figure}

Fig.\ref{fig:dustcrit} represents the results of all the models investigated so far in
a $Z$ -- $f_{\rm dep}$ plane. Empty symbols indicate no fragmentation and filled 
symbols represent models where fragmentation conditions are met. The solid line 
marks the border between the no-fragmentation and fragmentation zones and corresponds to
a minimum dust-to-gas ratio of only ${\cal D}_{\rm cr} = 4.4 \times 10^{-9}$. At any
given metallicity, low-mass fragmentation can be activated when ${\cal D} \ge {\cal D}_{\rm cr}$ or
$f_{\rm dep} \ge {\cal D}_{\rm cr}/Z$. 

A simple argument for the origin of ${\cal D}_{\rm cr}$ runs as follows.
The second high-density dip in the thermal evolution of collapsing gas clouds
is activated when the energy transfer rate between gas and dust is more
efficient than the compressional heating rate due to gravitational contraction,
that is,

\begin{equation}
H_{\rm gr}/\rho > -p  \,\frac{d}{dt} \,\frac{1}{\rho}. 
\label{eq:dustdip}
\end{equation}
\noindent 
Using eqs.~(\ref{eq:heat})-(\ref{eq:cross}) for the left-hand
term and eqs.~(\ref{eq:gas})-(\ref{eq:compression}) for the compressional
heating rate, eq.(\ref{eq:dustdip}) leads to a condition on the dust-to-gas
ratio times the cross section per unit dust mass,

\begin{equation}
S {\cal D} > 1.4 \times 10^{-3} \mbox{cm}^2/\mbox{gr} \left[\frac{T}{10^3 \mbox{K}}\right]^{-1/2} \left[\frac{n_{\rm H}}{10^{12} \mbox{cm}^{-3}}\right]^{-1/2},
\label{eq:sdust}
\end{equation}
\noindent
where we have assumed that at the onset of dust cooling, before gas and dust become thermally coupled, $T_{\rm gr} << T$.
The above relation has been evaluated at reference gas temperature and density of $T \sim 10^3$~K and $n_{\rm H} \sim 10^{12}$cm$^{-3}$. 
Indeed, the sharp temperature drop, which activates fragmentation, can be achieved only if dust cooling exceeds the 
compressional heating rate when $n_{\rm H} < 10^{12}$~cm$^{-3}$ and H$_2$ formation heating keeps the gas temperature at $\sim 10^3$~K. 
Beyond this density, H$_2$ becomes fully molecular, chemical heating is no longer effective, and dust--cooling causes only a mild
temperature decrease.
It follows from eq.~(\ref{eq:sdust}) that the critical dust-to-gas ratio is inversely proportional to the total grain cross section
per unit mass of dust. As it can be inferred from the last column of Table \ref{table:ejectacomp}, differences in $S$ 
among different reverse shock models for a given SN progenitor are less than $15\%$. 
Taking $S$ to vary in the range $2.22 \le S/10^5 \mbox{cm}^2/\mbox{gr} \le 5.37$, the
condition on the dust-to-gas ratio reads,
\begin{equation}
{\cal D} > [2.6 - 6.3] \times 10^{-9} \left[\frac{T}{10^3 \mbox{K}}\right]^{-1/2} \left[\frac{n_{\rm H}}{10^{12} \mbox{cm}^{-3}}\right]^{-1/2},
\label{eq:dustcrit}
\end{equation}
\noindent
consistent with value of ${\cal D}_{\rm cr}$ inferred above and drawn in the figure. 

It is important to stress that this simple analytic argument is not able to capture 
all the important details of the evolution which may activate dust-induced fragmentation.
In particular, the condition that $\gamma < 0.8$ during the dust-induced cooling phase
depends also on the grain opacity: if we increase the dust opacity for fixed surface area,
the dust-gas coupling temperature becomes lower, leading to a deeper dip in the equation
of state.

In addition, the fragmentation criteria that we follow 
($\gamma > 0.97$ after a phase of cooling where $\gamma < 0.8$) have been established 
through a comparison with numerical studies (Tsuribe \& Omukai 2006; Omukai et al. 2010).
It is clear that the process of fragmentation does not depend only on the gas thermal evolution,
being influenced also by the detailed hydrodynamical properties of the clouds, such as the initial
amplitude of fluctuations, rotation and turbulence. Thus, our study can provide a guideline for 
more sophisticated numerical simulations.

\section{Discussion} 
Although we have considered only star--forming regions in the first mini--halos, 
metals and dust rapidly enrich more massive halos, with masses $M \sim 10^8 M_{\odot}$ at $z \sim 10$,
the so--called first galaxies (Bromm \& Yoshida 2011). When the gas is accreted onto these halos, it is
shock--heated at the virial radius of the parent dark matter
halo to virial temperatures $T_{\rm vir} \sim 10^4$~K; 
then it cools via the atomic--hydrogen Lyman--$\alpha$ line, falling inward and attaining increasingly larger densities.
In Schneider et al. (2002), we investigate the dependence of thermal evolution of collapsing gas clouds on the initial
virial temperature (halo mass). For metal--free gas, the evolution with temperature of the ionization fraction and of the
fraction of molecular hydrogen is independent of the halo virial temperature. If the gas is fully ionized, molecules form 
through nonequilibrium recombination, leading to overall 
fractions that are much higher than in the expanding homogeneous universe (Uehara \& Inutsuka 2000). 
As a consequence of enhanced H$_2$ and HD fractions, the gas rapidly cools to the CMB temperature and $\sim 10 M_{\odot}$ fragments are formed 
(Johnson \& Bromm 2006). 

In the presence of a strong LW background, the gas in the first galaxies remains free of H$_2$ (Omukai 2001), it avoids fragmentation
and leads to the formation of a single super--massive star with mass
as high as $\sim [10^5 - 10^6] M_{\odot}$,  which then evolves into a super--massive black hole (Bromm \& Loeb 2003b).   
Omukai, Schneider \& Haiman (2008) find that, even in these systems, fragmentation is inevitable when the
gas is pre--enriched above a critical metallicity, whose value is between  
$3 \times 10^{-2} Z_{\odot}$  (in the absence of dust) and as low as $5\times 10^{-6} Z_{\odot}$ 
(with a dust--to--gas mass ratio of about $0.01 Z/Z_{\odot}$).
 
Recently, Dekel \& Birnboim (2006) and Dekel et al. (2009) have proposed an alternative accretion mode, 
the so--called `cold accretion': gas falls in the halos along cosmic filaments and it is deposited 
much closer to the center of the galaxy, at much
higher densities. In these conditions, the fragmentation properties of the gas are likely to be modified. 
Safranek-Shrader, Bromm \& Milosavljevi{\'c} (2010) find that H$_2$ and HD dominate gas cooling at $T < 10^3$~K 
and metallicity has no effect on fragmentation. In the presence of a strong LW background which photo--dissociates
H$_2$, metal--line cooling becomes important and, when the metallicity is $Z \sim 10^{-4} Z_{\odot}$, $\sim 3 M_{\odot}$
fragments are formed. While this appears an interesting possibility to form low--mass stars, the final fragmentation
mass depends sensitively on the adopted initial gas density. Thus, a larger set of initial conditions needs to
be investigated to draw robust conclusions.    

Finally, it is important to mention the effects of the CMB radiation field on the thermal evolution of collapsing gas clouds,
which have not been included in the present study. In general, the presence of the CMB radiation field is important at high
redshifts and metallicities (Smith et al. 2009), inhibiting fragmentation. Schneider \& Omukai (2010) find that in the
absence of dust grains, the fragmentation mass scales at $z > 10$ are always $\ge$ 100s of $M_{\odot}$,
independent of the gas initial metallicity. When $Z < 10^{-2} Z_{\odot}$, 
dust--cooling remains relatively insensitive to the presence of the CMB, and 
sub--solar mass fragments are formed at any redshift. 
It is only when $Z > 10^{-2} Z_{\odot}$ and $z > 10$ that the heating of dust grains by the CMB 
starts to inhibit the formation of sub--solar mass fragments (Schneider \& Omukai 2010).
Thus, the conclusions of the present study remain valid in all but these extreme conditions.

\section{Conclusions}

In this study, we have explored the conditions which enable the formation of solar and sub--solar
mass fragments during the collapse of pre--enriched gas clouds within the first mini--halos at $z \sim 20-30$.

By carefully analysing the thermal and chemical evolution of star--forming clouds, we find that in the
absence of dust grains, gas fragmentation occurs at densities $n_{\rm H} \sim [10^4-10^5]$~cm$^{-3}$
when the metallicity exceeds $Z \sim 10^{-4} Z_{\odot}$. The resulting fragmentation masses,
identified by the minimum Jeans mass at the fragmentation epoch, are $\ge 10 M_{\odot}$.
The inclusion of Fe and Si cooling (which can dominate metal--line emission
in some density ranges) does not affect the thermal evolution as this is dominated by
molecular (mostly OH, H$_2$O and CO) cooling even for metallicities as large as
$Z = 10^{-2} Z_{\odot}$.

The presence of dust is the key driver for the formation of solar--mass stars.
We have explored the fragmentation properties of collapsing clouds pre--enriched
with dust and metals synthesized by core-collapse SNe. Focusing on three representative
SN progenitors (a $Z=0$ star with $20 M_{\odot}$ and two $Z=10^{-4} Z_{\odot}$ stars
with $20 M_{\odot}$ and $35 M_{\odot}$), we compute the mass of dust and metals
(hence the depletion factor), grain composition and size distribution using the
model by Bianchi \& Schneider (2007). To determine the minimum dust depletion factor
required to activate dust--driven fragmentation, we start from the above initial conditions
and consider the effects of SN reverse shocks of increasing strength: these
reduce the depletion factors, alter the shape of the grain
size distribution function and modify the relative abundances of grain species and
of metal species in the gas phase. Following this approach, we can control all these
effects in a self--consistent way and explore the implications
for the fragmentation properties of pre--enriched star forming clouds.

We find that the lowest metallicity at which fragmentation occurs is $Z=10^{-6} Z_{\odot}$
for gas pre--enriched by the explosion of a $20 M_{\odot}$ primordial SN and/or by a $35 M_{\odot}$,
$Z=10^{-4} Z_{\odot}$ SN;
it is $\sim$ 1 dex larger, when the gas is pre--enriched by a $Z = 10^{-4} Z_{\odot}$, 20~$M_{\odot}$ SN.
In both cases, however, cloud fragmentation depends on the
depletion factor and it is suppressed when the
reverse shock leads to a too large destruction of dust
grains.

These features are all consistent with the idea that there
is a minimum dust--to--gas ratio, ${\cal D}_{\rm cr}$,
above which fragmentation is activated. By requiring that the
energy transfer rate between gas and dust is larger than the
compressional heating rate, we derive a simple analytic
expression for ${\cal D}_{\rm cr}$ which depends on the
total grain cross--section per unit mass of dust; for grain
composition and properties explored in the present study, this
critical dust--to--gas ratio is ${\cal D}_{\rm cr} = [2.6 - 6.3] \times 10^{-9}$.
When the dust--to--gas ratio of star forming clouds exceeds this value,
the fragmentation masses range between $0.01 M_{\odot}$ and $1 M_{\odot}$,
thus enabling the formation of the first low--mass stars.

According to this criterium, the recently observed Galactic Halo star with
a total metallicity of $Z = 4.5 \times 10^{-5} Z_{\odot}$ (Caffau et al. 2011)
requires a minimum dust depletion of $f_{\rm dep} \ge (4 - 9) \times 10^{-3}$ 
in the parent gas cloud, consistent with the values predicted for ordinary
core--collapse SN investigated in the present study.
   
\section*{Acknowledgments}
We greatfully acknowledge all members of the DAVID collaboration (www.arcetri.astro.it/david) for inspiration
and stimulating discussion. 
The present work is supported in part by the Grants-in-Aid
by the Ministry of Education, Science and Culture of Japan
(2168407, 21244021:KO).

\label{lastpage}
\end{document}